\documentclass[structabstract]{aa}  
\usepackage[pdftex]{graphicx}
\usepackage{txfonts}
 \newcommand{\hii}{\relax \ifmmode {\mbox H\,{\scshape ii}}\else H\,{\scshape ii}\fi}
\newcommand{\mi}{\relax \ifmmode {\mu{\mbox m}}\else $\mu$m\fi}
\newcommand{\ha}{\relax \ifmmode {\mbox H}\alpha\else H$\alpha$\fi}
\newcommand{\hb}{\relax \ifmmode {\mbox H}\beta\else H$\beta$\fi}
\newcommand{\hg}{\relax \ifmmode {\mbox H}\beta\else H$\gamma$\fi}
\newcommand{\hd}{\relax \ifmmode {\mbox H}\beta\else H$\delta$\fi}

\newcommand{\sii}{\relax \ifmmode {\mbox S\,{\scshape ii}}\else S\,{\scshape ii}\fi}
\newcommand{\siii}{\relax \ifmmode {\mbox S\,{\scshape iii}}\else S\,{\scshape iii}\fi}
\newcommand{\nii}{\relax \ifmmode {\mbox N\,{\scshape ii}}\else N\,{\scshape ii}\fi}
\newcommand{\oii}{\relax \ifmmode {\mbox O\,{\scshape ii}}\else O\,{\scshape ii}\fi}
\newcommand{\oiii}{\relax \ifmmode {\mbox O\,{\scshape iii}}\else O\,{\scshape iii}\fi}
\newcommand{\neiii}{\relax \ifmmode {\mbox Ne\,{\scshape iii}}\else Ne\,{\scshape iii}\fi}
 
 \newcommand{\rdostres}{\relax \ifmmode {\,\mbox{R}}_{\rm 23}\else \,\mbox{R}$_{\rm 23}$\fi}

\begin{document}
   \title{The cosmic evolution of oxygen and nitrogen abundances in star-forming galaxies over the last 10 Gyrs}
%   \title{The cosmic evolution of metallicity and nitrogen-based relations of star-forming galaxies in the 20k zCOSMOS sample}

\author{
E.~P\'erez-Montero\inst{1,2,3}
\and
T.~Contini\inst{1,2}
\and
F.~Lamareille\inst{1,2}
\and  %%%%%%%%%%%%%%%%%%%%%%%%%%%%%%%%%%%%%%%% CORE A+ 
C.~Maier\inst{4,15}
\and
C.~M.~Carollo\inst{4}
\and
J.-P.~Kneib\inst{5}
\and
O.~Le~F\`{e}vre\inst{5}
\and
S.~Lilly\inst{4}
\and
V.~Mainieri\inst{6}
\and
A.~Renzini\inst{7}
\and
M.~Scodeggio\inst{8}
\and
G.~Zamorani\inst{9}
\and  %%%%%%%%%%%%%%%%%%%%%%%%%%%%%%%%%%%%%%%% CORE A
S.~Bardelli\inst{9}
\and
M.~Bolzonella\inst{9}
\and 
A.~Bongiorno\inst{10}
\and
K.~Caputi\inst{20}
\and
O.~Cucciati\inst{12}
\and
S.~de~la~Torre\inst{11}
\and
L.~de~Ravel\inst{11}
\and
P.~Franzetti\inst{8}
\and
B.~Garilli\inst{8,5}
\and
A.~Iovino\inst{13}
\and
P.~Kampczyk\inst{4}
\and 
C.~Knobel\inst{4}
\and
K.~Kova\v{c}\inst{4,14}
\and
J.-F.~Le~Borgne\inst{1,2}
\and
V.~Le~Brun\inst{5}
 \and
M.~Mignoli\inst{9}
\and 
R.~Pell\`o\inst{1,2}
\and
Y.~Peng\inst{4}
\and
V.~Presotto \inst{16,13}
\and
E.~Ricciardelli\inst{17}
\and
J.~D.~Silverman\inst{18}
\and
M.~Tanaka\inst{18}
\and 
L.~A.~M.~Tasca\inst{5}
\and
L.~Tresse\inst{5}
\and
D.~Vergani\inst{9,19}
\and
E.~Zucca\inst{9}
}

 \offprints{E. P\'erez-Montero \email{epm@iaa.es}}

  \institute{
{Institut de Recherche en Astrophysique et Plan{\'e}tologie, CNRS, 14, avenue Edouard Belin, F-31400 Toulouse, France} % 1
\and
{IRAP, Universit{\'e} de Toulouse, UPS-OMP, Toulouse, France} % 2
\and
{Instituto de Astrofis\`ica de Andaluc\`ia, CSIC, Apartado de correos 3004, 18080 Granada, Spain}  % 3
\and
{Institute of Astronomy, ETH Zurich, CH-8093, Z\"urich, Switzerland}   % 4
\and
{Laboratoire d'Astrophysique de Marseille, CNRS-Universit{\'e} d'Aix-Marseille, 38 rue Frederic Joliot Curie, F-13388 Marseille, France}  % 5
\and
{European Southern Observatory, Karl-Schwarzschild-Strasse 2, Garching b. Muenchen, D-85748, Germany}  % 6
\and
{Dipartimento di Astronomia, Universit\`a di Padova, vicolo Osservatorio 3, I-35122 Padova, Italy}  % 7
\and
{INAF - IASF Milano, Via Bassini 15, I-20133, Milano, Italy}   % 8
\and
{INAF - Osservatorio Astronomico di Bologna, via Ranzani 1, I-40127 Bologna, Italy}  % 9
\and 
{Max-Planck-Institut f\"ur extraterrestrische Physik, D-84571 Garching b. Muenchen, D-85748, Germany}  % 10
\and 
{SUPA Institute for Astronomy, The University of Edinburgh, Royal Observatory, Edinburgh, EH9 3HJ}  % 11
\and
{INAF - Osservatorio Astronomico di Trieste, Via Tiepolo, 11, I-34143 TRIESTE, Italy}  % 12
\and
{INAF - Osservatorio Astronomico di Brera, Via Brera, 28, I-20159 Milano, Italy}  % 13
\and
{MPA - Max Planck Institut f\"ur Astrophysik, Karl-Schwarzschild-Str. 1,  85741 Garching, Germany}  % 14
\and 
{University of Vienna, Department of Astronomy, Tuerkenschanzstrasse 17, 1180 Vienna, Austria} %15
\and
{Universit\'a degli Studi dell'Insubria, Via Valleggio 11, 22100 Como, Italy} % 16
\and 
{Instituto de Astrof\`sica de Canarias, V\'ia Lactea s/n, E-38200 La Laguna, Tenerife, Spain} % 17
\and
{IPMU, Institute for the Physics and Mathematics of the Universe, 5-1-5 Kashiwanoha, Kashiwa, 277-8583, Japan}  % 18
\and
{INAF-IASFBO, Via P. Gobetti 101, I-40129, Bologna, Italy}  % 19
\and
{Kapteyn Astronomical Institute, University of Groningen, 9700 AV Groningen,
The Netherlands} % 20
  }

   \date{}  

 \abstract
% Context (optional)
{}
% Aims (mandatory)
{The chemical evolution of galaxies on a cosmological timescale is still a matter of debate despite the
increasing number of available data provided by spectroscopic surveys of star-forming galaxies at different
redshifts. The fundamental relations involving metallicity, such as the mass-metallicity 
relation ($MZR$) or the fundamental-metallicity relation, give controversial results about
the reality of evolution of the chemical content of galaxies at a given stellar mass. 
In this work we shed some light on this issue using the
completeness reached by the 20k bright sample of the zCOSMOS survey and using for
the first time the nitrogen-to-oxygen ratio (N/O) as a 
tracer of the gas phase chemical evolution of galaxies that is independent of
the star-formation rate.}
% Methods (mandatory)
{Emission-line galaxies both in the SDSS and 20k zCOSMOS bright survey were
used to study the evolution from the local Universe 
of the  $MZR$ up to a redshift of $\sim$ 1.32, and the relation between stellar 
mass and N/O ($MNOR$) 
up to a redshift of $\sim$ 0.42  using the N2S2 parameter. All the physical properties derived 
from stellar continuum and gas emission-lines, including stellar mass, star formation rates, 
metallicity and N/O, were calculated in a self-consistent way over the full redshift range.}
% Results (mandatory)
{We confirm the trend to find lower metallicities in galaxies of a given stellar mass 
in a younger Universe.
% in all redshift and stellar mass bins studied with a significant level of completeness and with a sufficiently large number of galaxies.
This trend is even observed when taking possible effects into account that are due to
the observed larger median star formation rates for galaxies at higher redshifts. 
We also find a significant evolution of the $MNOR$ up to $z\sim 0.4$.
%study the evolution of the $MNOR$ finding a significant evolution of this ratio in the studied redshift range. 
Taking the slope of the O/H vs. N/O relation into account for the secondary-nitrogen production regime, the observed evolution 
of the $MNOR$ is consistent
with the trends found for both the $MZR$ and its equivalent relation using new
expressions to reduce its dependence on star-formation rate.}
% Conclusions (optional)
{} 
% 5 {} token are mandatory


\titlerunning {Cosmic evolution of galaxy metal content over the last 10 Gyrs}        
\authorrunning{E. P\'erez-Montero et al.}
  \maketitle
 
%
%________________________________________________________________

\section{Introduction}
Chemical abundances provide important clues to the evolutionary history 
of galaxies along cosmic time since they are the result of the joint action 
of several physical processes such as supernova feedback, gas inflow/outflow, 
mergers, and interactions, Even if it is not easy to disentagle 
the effect of each individual process, it may be possible to constrain galaxy 
formation theories by studying how chemical abundances for galaxies of different 
masses and/or in different environments evolve as a function of cosmic 
epoch. Indeed, as cosmological time progresses, galaxy evolution models predict 
that both the mean metallicity and stellar mass of galaxies increase with age 
as galaxies undergo chemical enrichment and grow through merging 
processes. At any given epoch, the accumulated history of star formation, 
gas inflows, and outflows, affects a galaxy mass and its metallicity. One therefore
expects these quantities to be correlated in some way and this correlation 
to provide crucial information about the physical processes that govern 
galaxy formation. 

First discovered for dwarf irregular galaxies (Lequeux et al. 1979), the relation between 
the mass (or luminosity) and the metallicity for different galaxy populations has been 
the subject of numerous studies (e.g. Skillman et al. 1989; Brodie \& Huchra 1991; 
Zaritsky et al. 1994; Garnett et al. 1997; Pilyugin \& Ferrini 2000; Contini et al. 2002). 
The mass--metallicity relation (hereafter $MZR$) is now well established in the local
universe, thanks to the works of Tremonti et al. (2004) based on SDSS and Lamareille et al. (2004)
based on 2dFGRS data, showing that the metallicity of galaxies 
tends to increase with their mass or luminosity. This trend has been shown to extend to 
much lower galaxy masses (Lee et al. 2006; Saviane et al. 2008), confirming the idea 
that a single mechanism may govern galaxy metallicities across five order of magnitude 
in stellar mass. However, a specific environment and/or high star formation rate can explain 
the observed offset of galaxy population with respect to the global $MZR$. For instance,  
nuclear inflows of metal-poor interstellar gas triggered by galaxy interactions can account for the
systematically lower central oxygen abundances observed in interacting galaxies (P\'erez et al. 2011; Torrey et al. 2012).

Hierarchical galaxy formation models that take the chemical 
evolution and feedback processes into account are able to reproduce the observed $MZR$ in 
the local universe (e.g. De Lucia et al. 2004; De Rossi et al. 2007; 
Finlator et al. 2007; Dav\'e et al. 2012). 
However these models rely on free parameters, such as 
feedback efficiency, which are not yet well constrained by observations. 
Alternative scenarios proposed to explain the $MZR$ include low star formation 
efficiency in low-mass galaxies caused by supernova feedback (Brooks et al. 2007) 
and a variable stellar initial mass function that is more top-heavy in galaxies with 
higher star formation rates, thereby producing higher metal yields (K\"oppen  et al. 2007). 

The evolution of the $MZR$ on cosmological timescales is now predicted by semi-analytic 
models of galaxy formation, which include chemical hydrodynamic simulations within 
the standard $\Lambda$-CDM framework (De Lucia et al. 2004; Dav\`e \& Oppenheimer 2007; 
Sakstein et al. 2011). 
Reliable observational estimates of the $MZR$ of galaxies at different epochs 
(hence different redshifts) may thus provide important constraints on galaxy evolution 
scenarios. Estimates of the mass--metallicity - or luminosity--metallicity
- relation of galaxies up to $z\sim1.5$ have been already derived but
have been limited, until recently, to small samples ({\em e.g.} Kobulnicky et al. 2003; Liang et al. 2004, 
Maier et al. 2004, 2005, 2006; Hammer et al. 2005; Savaglio et al. 2005; Lamareille et al. 2006; 
Liu et al. 2008; Queyrel et al. 2009; Yabe et al. 2012). Recent studies of the $MZR$ (Lamareille et al. 2009; 
P\'erez-Montero et al. 2009; Cowie \& Barger 2008; Moustakas et al. 2011; 
Zahid et al. 2011) have been performed on larger samples 
($>1000$ galaxies) thanks to the large and deep spectroscopic surveys (VVDS, DEEP2, 
GOODS, AGES, etc) devoted to the formation and evolution of galaxies on cosmological timescales. 
The general observational result of these studies is that the $MZR$ evolves with redshift, in the sense that, 
on average and for a given stellar mass, high-redshift galaxies are characterised by lower 
metallicities. Whether the $MZR$ also evolves in terms of shape/slope 
is still matter of debate. Indeed, Savaglio et al. (2005) conclude that there is a steeper slope in 
the distant universe, whereas Lamareille et al. (2009) and P\'erez-Montero et al. (2009) 
found flattening of the $MZR$ at redshifts $0.7 < z < 1.0$.  At higher redshifts Erb et al. (2006) 
derived a $MZR$ at $z\sim 2$ lowered by $0.3$ dex in metallicity compared with 
the local estimate, a trend that could extend up to $z\sim 3-4$ (Maiolino et al. 2008; Mannucci et al. 2009). 

However, the redshift evolution of the gas-phase metallicity in galaxies has been questioned recently 
by Mannucci et al. (2010) and Lara-L\'opez et al. (2010). They discovered that metallicity depends not only on the stellar mass, but also 
on the star formation rate (SFR): for a given stellar mass, galaxies with higher SFR systematically show 
lower metallicities. This is the so-called fundamental metallicity relation ($FMR$), i.e., a tight relation between 
stellar mass, gas-phase metallicity, and SFR. Local SDSS galaxies show very small residuals ($\sim 0.05$ dex) 
around this relation (Yates et al. 2012; Brisbin \& Harwit 2012). According to Mannucci et al. (2010), the $FMR$ 
does not evolve with redshift up to $z\sim 2.5$.  This result was later confirmed by Cresci et al. (2012)
by studying the zCOSMOS 10k bright sample up to a redshift $\approx$ 0.8.
This would suggest that the observed evolution of the $MZR$ 
is due to selection effects and to the well-established increase in the average SFR with redshift (Noeske et al. 
2007; Elbaz et al. 2007; Daddi et al. 2007). 

The study of other indicators of the chemical content of galaxies and their relation
with stellar mass and SFR have not been sufficiently explored so far so can shed
some light on the issue of the evolution of metallicity with cosmic age. This is the case
of the nitrogen-to-oxygen abundance ratio (N/O), which offers some clear advantages over
studies of the oxygen abundance alone. On one hand, N is produced mainly by
low- and intermediate-mass stars (Henry et al. 2000) in contrast to O, which is produced
only by massive stars. This makes N/O a suitable indicator for the chemical
evolution of single starbursts (Edmunds \& Pagel, 1978; Pilyugin et al.  2003). On the other hand, since 
N is a secondary element ({\em i.e.} its yield depends on the previous amount of
carbon and oxygen in the stars) for the majority of the metallicity range, its relation
with a primary element, such as O, is relatively independent of the chemodynamical effects,
such as outflows or inflows (Edmunds, 1990, K\"oppen \& Hensler, 2005). 
These effects are the basis of the relation between metallicity
and SFR, so the relation between N/O and stellar mass (hereafter $MNOR$) can be 
fundamental for studying the chemical evolution of galaxies without worrying about the selection effects
of the sampled galaxies at high $z$.
The $MNOR$ for galaxies in the local universe has already been investigated by 
P\'erez-Montero \& Contini (2009, hereafter PMC09) who found an increase in N/O with 
stellar mass, since most of the sampled SDSS galaxies lie in the metallicity range for
the production of secondary N. This relation was used by Amor\'\i n et al. (2010) to study the chemical
evolution of compact galaxies with very high specific star formation rates (the {\em green pea}
galaxies). Later, Thuan et al. (2010) explored the evolution of the relation between stellar mass and N/H
up to $z$ = 0.3, exploiting the fact that the production rate of N is faster than for O
and its evolution is thus less sensitive to calibration uncertainties.

In this work we investigate both the $MZR$ with its relation to SFR and the $MNOR$ in
the Local Universe and their evolutions with
cosmic age. This study is based on the SDSS sample for the local universe and the zCOSMOS 20k 
bright sample for high redshifts (up to a $z \approx 1.32$ for the $MZR$ and $z \approx 0.42$
for the $MNOR$). 

This paper is organised as follows.  In section 2 we describe the SDSS and zCOSMOS samples
of the star-forming galaxies used to perform our study of the evolution of both the $MZR$ and 
the $MNOR$. 
The selection of star-forming galaxies and
the derivation of physical properties, including stellar mass, star formation rate, 
oxygen abundance, and the N/O abundance ratio are described in section 3. In section 4 
we give our results and discuss the evolution with cosmic age of the $MZR$, with a correction
to take selection effects due to SFR into account, and of the $MNOR$. Finally,
in section 5 we summarise our work and give our conclusions.

Throughout this paper we normalise the derived stellar masses and the absolute magnitudes 
with the standard $\Lambda$-CDM cosmology, {\em i.e.}, $h$ = 0.7, $\Omega_m$ = 0.3 and 
$\Omega_\Lambda$ = 0.7 (Spergel et al., 2003).

\section{Sample selection}

\subsection{The parent SDSS local sample}
\label{sdss}

Ss reference sample for the local universe, we used the Data Release 7 of
the Sloan Digital Sky Survey (hereafter SDSS), taking the emission-line measurements from the 
Max Planck Institute for Astrophysics-Johns Hopkins University (MPA-JHU)
catalogue\footnote{Available at http://www.mpa-garching.mpg.de/SDSS/}.
The emission-line measurement of this sample is described in Brinchmann et al. (2004).
From the whole sample of galaxies, we removed duplicated objects and
all those emission-line galaxies whose \hb, [\oii] $\lambda\lambda$ 3727, 3729, 
[\oiii] $\lambda\lambda$ 4959, 5007, \ha, [\nii] $\lambda$ 6584, and 
[\sii] $\lambda\lambda$ 6717,6731 have 
signal-to-noise ratio lower than 2.  The choice for this low S/N was
motivated by the search for a coincidence between the mass-metallicity 
relation of the SDSS sample derived in this work and the zCOSMOS subsample at low $z$, 
avoiding biases in the sample selection for the derivation of metallicity,
as discussed below.
The selected sample comprises a total of 299\,479 galaxies
in the $0.02  < z < 0.20$ range. but with a mean value
$z = 0.068$. The lower redshift limit is caused by the lower
limit of the spectral coverage of the SDSS sample (3800 \AA), which prevents
the detection of the [\oii] $\lambda$ 3727 emission-line at $z < 0.02$.
This lower limit also prevents the sample from being contaminated by 
nearby objects with an inaccurate determination of their integrated 
properties due to the limited size of the fibre in the SDSS. 
as they were not totally covered by the 3\arcsec\ fibre of SDSS.
All collected emission lines were extinction-corrected using
the law by Cardelli et al. (1989) and the reddening constants 
from the Balmer decrement between H$\alpha$ and H$\beta$ as compared
to the theoretical value for mean nebular conditions given by
Storey \& Hummer (1995).

\subsection{The parent zCOSMOS 20k sample}

The COSMOS survey is a large HST-ACS survey, with $I$-band exposures
down to $I_{\rm AB}=28$ on a field of $\sim 2$ deg$^2$ (Scoville et al. 2007). 
The COSMOS field has been the object of extensive multiwavelength 
ground- and space-based observations spanning the entire spectrum: X-ray, 
UV, optical/NIR, mid-infrared, mm/submillimetre and radio, providing fluxes 
measured over 30 photometric bands (Hasinger et al. 2007; Taniguchi et al. 2007; 
Capak et al. 2007, Lilly et al. 2007; Sanders et al. 2007;  Bertoldi et al. 2007; 
Schinnerer et al. 2007; Koekemoer et al. 2007; McCracken et al. 2010).

The main spectroscopic
follow-up survey undertaken in the COSMOS field, zCOSMOS (Lilly et al. 2007), used 600
hours of ESO observing time with the VIMOS multi-object spectrograph
(Le F\`evre et al. 2003) mounted on the Melipal 8m-telescope
of the ESO-VLT. The zCOSMOS spectroscopic survey
consists of two parts: zCOSMOS-bright and
zCOSMOS-deep. The zCOSMOS-deep targets $\sim 10\,000$ galaxies within
the central 1 deg$^2$ of the COSMOS field, selected through colour 
criteria to have $1.4 \la z\la 3.0$. The zCOSMOS-bright is purely
magnitude-limited and covers the whole area of 1.7 deg$^2$ of the
COSMOS field. It provides redshifts for $\sim 20\,000$ galaxies down to
$I_{\rm AB} \leq 22.5$ as measured from the HST-ACS imaging.  The
success rate in redshift measurements is very high, 95\% in the
redshift range $0.5 < z < 0.8$, and the velocity accuracy is $\sim
100$ km/s (Lilly et al. 2009). Each observed object has been assigned
a flag according to the reliability of its measured redshift. Classes
3.x, 4.x redshifts, plus Classes 1.5, 2.4, 2.5, 9.3, and 9.5 are
considered a secure set, with an overall reliability of 99\% 
(see for details Lilly et al. 2009).

For this work, we used the zCOSMOS-bright survey final
release: the so-called 20k sample for about $20\,000$ galaxies with 
$z\leq 2$ and secure redshifts according to the above flag 
classification ($18\,206$ objects in total, irrespective of redshift and 
including stars). Observations for the zCOSMOS-bright survey were 
acquired with the medium-resolution ($R = 600$) grism of VIMOS, 
providing spectra over the $5550-9650$\AA\  wavelength range. 
The total integration time was set to one hour to secure redshifts with 
a high success rate. The spectral range of observations enables 
following important diagnostic emission lines to compute metallicity up to redshift
$z\sim 1.5$. The observations were acquired with a seeing lower that 1.2\arcsec.

\subsubsection{Emission-line measurement}
\label{meas}

Spectroscopic measurements (emission and absorption lines
strength) in zCOSMOS were performed through the automated pipeline {\it platefit-vimos} 
(Lamareille et al., in preparation) similar to the one performed
on SDSS ({\em e.g.} Tremonti et al. 2004) and VVDS spectra
(Lamareille et al. 2009). This routine processes spectra in two
steps. The stellar component of galaxy spectra is fitted as a combination
of 30 single stellar population templates, with different
ages and metallicities, from the library of Bruzual \& Charlot
(2003) resampled to the velocity resolution of zCOSMOS spectra.
After removal of the stellar component, emission lines are
fitted together as a single nebular spectrum made of a sum of
Gaussians at specified wavelengths. Further details can be found
in Lamareille et al. (2006, 2009).

For this work we selected those zCOSMOS emission-line galaxies
with an S/N higher than two for the involved emission-lines in
each redshift regime, which leaves a total of 5\,331 objects.

\section{Physical properties}

\subsection{Selection of star-forming galaxies}
\label{agnsel}

%______________________________________________ 
 \begin{figure*}
\begin{minipage}{180mm}
  \centerline{
   \includegraphics[width=8cm,clip=]{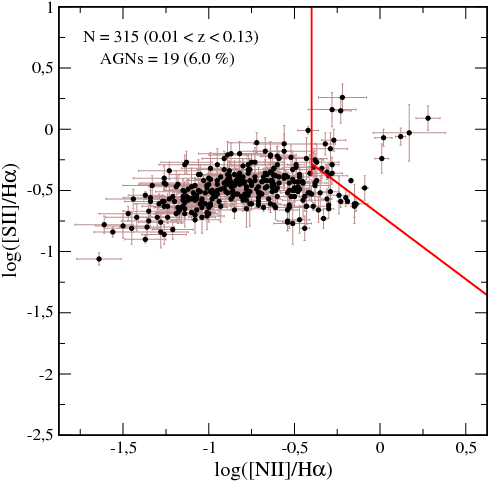}
   \includegraphics[width=8cm,clip=]{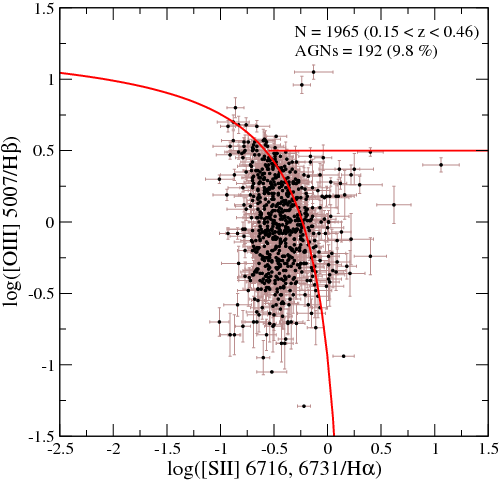}}
   \centerline{
   \includegraphics[width=8cm,clip=]{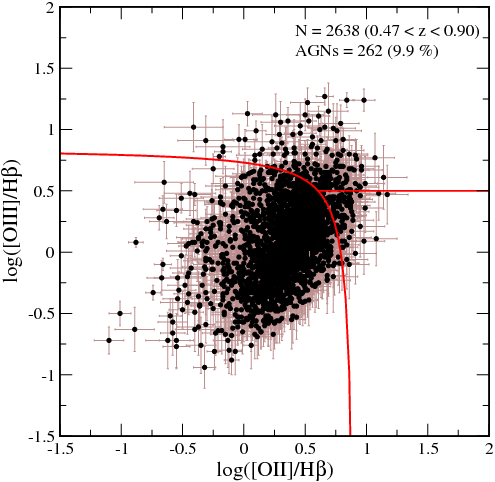}
   \includegraphics[width=8cm,clip=]{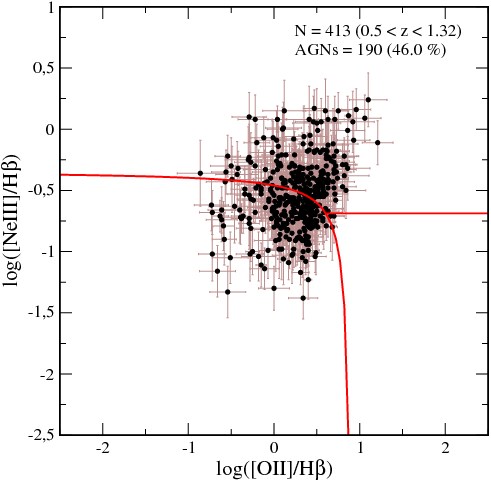}}
      \caption{Diagnostic diagrams used to select star-forming objects in
      the studied sample in each redshift range. From left to right and from up to down:
      [\nii]/\ha\ vs. [\sii]/\ha, [\nii]/\ha\ vs. [\oiii]/\hb, 
      [\oii]/\hb\ vs. [\oiii]/\hb, and [\oii]/\hb\ vs. [\neiii]\hb.}
	\label{diagn}
  
\end{minipage}
  \end{figure*}
%____________________________________

We selected star-forming galaxies from both the parent SDSS and 20k zCOSMOS
samples by means of several empirical diagnostic diagrams
based on bright emission-line ratios. We identified narrow-line
AGNs (Seyfert 2 and LINERs), taking into account that the
broad-line AGNs were already removed. 
The diagnostic diagrams used for the final classification are
not the same for all the galaxies, as the available emission lines depend on the 
considered redshift range. Indeed, the strategy of the zCOSMOS survey, with
a fixed 5550-9650 \AA\ wavelength range, implies that the set of
bright emission-lines used in the diagnostic diagrams changes
with redshift. This is also true for the oxygen abundance
determination, which is based on the same sets of emission lines
(see section \ref{meta}).

For the 315 zCOSMOS emission-line galaxies in the redshift 
range 0.01 $<$ $z$ $<$ 0.13, we used
the \ha\ classification proposed by Lamareille (2007) based on
[\nii], [\sii], and \ha\ emission-line ratios and independent of the reddening. 
According to this classification, objects are classified as narrow-line AGNs if
\[\log(\textrm{[\sii]/\ha}) \geq -0.3, \textrm{  and}\] 
\begin{equation}
\log([\textrm{\nii}]/\textrm{\ha}) > -0.4 
\end{equation}
\noindent or
\begin{equation}
\log([\textrm{\nii}/\textrm{\ha}) > -1.05 \cdot \log([\textrm{\sii}]/\textrm{\ha}) .
\end{equation}
This gives a total of 19 AGNs (6\%), which are rejected from the
sample to be analysed. The [\nii]/\ha\ vs. [\sii]/\ha\ plot for the sample of 
zCOSMOS 20k emission-line galaxies in this redshift regime is shown in upper left hand panel of Fig. \ref{diagn}.

For both the redshift range 0.15 $<$ $z$ $<$ 0.44 in the zCOSMOS sample and for the
SDSS parent sample, we used the well known and commonly used diagnostic diagrams (e.g. Baldwin
et al. 1981; Veilleux \& Osterbrock 1987) based on the
line ratios  [\oiii]/\hb\ and [\sii]/\ha.
In this diagnostic diagram, all points below the
separation curve defined by Kewley et al. (2001) can be considered
as purely star-forming objects. 
Since our analysis for both oxygen and nitrogen abundances are based mainly on [\nii]
emission-line we preferred not to use the diagram that depends on the [\nii]/\ha\ for
two reasons: 
i) this diagnostic imposes an upper limit on the N2 = log([\nii]/\ha) parameter, and hence
for the upper metallicity and ii) according to PMC09,
both the {\em composite} ({\em i.e} a region in the diagram where are objects
with a mixture of ionising sources, Kewley et al. 2006) and the AGN region can
be contaminated with pure star-forming objects with very high N/O.
The total number of zCOSMOS galaxies classified as narrow-line AGN in this 
redshift range is 188, which is 9.6\% of the sample. The right 

hand upper panel of Fig. \ref{diagn} shows 
the sample, along with the curve of separation. The limit log([\oiii]/\hb) = 0.5
shows the separation between Sey2 and LINER galaxies.
Regarding SDSS, a total of 44\,729 objects were classified as narrow-line AGNs
(17.5 \% of the sample) and 254\,750 objects were selected as star-forming galaxies
for the analysis of the fundamental relations in the local universe.

At higher redshift (0.47 $<$ $z$ $<$ 0.9), the \ha, [\nii] $\lambda$ 6584, and
[\sii] $\lambda\lambda$ 6717,6731 emission lines are not visible anymore in the
observed wavelength range of zCOSMOS spectra. In this redshift
range, we instead used the blue diagnostic diagrams,
as defined by Lamareille et al. (2004), involving the [\oiii]/\hb\
and [\oii]/\hb\ emission-line ratios. Unlike the diagnostic
diagrams used for lower redshift, this method does depend on reddening due
to the long wavelength baseline between [\oii] $\lambda$ 3727 \AA\ and
\hb. One way to minimise this dependence is to use the ratio of equivalent
widths instead of fluxes as proposed by Kobulnicky et al. (2003), but this
method must be corrected owing to the different shape of the underlying 
stellar continuum in each galaxy. To do so, we took the expression proposed by
Liang et al. (2007), who propose the use of a multiplicative factor $\alpha$ to the
ratio of equivalent widths of [\oii] and \hb\ as a function of the
spectral stellar break at 4000 \AA. In our case, this break was automatically
measured in all the zCOSMOS 20k spectra at this $z$ by {\em platefit-vimos} in the same way as described
in section \ref{meas}. By considering the separation curve proposed
by Marocco et al. (2011), plotted in the left hand lower panel of Fig. \ref{diagn}, 
we classified 262 galaxies (9.9\%) as narrow-line AGNs, and thus we rejected them.

Finally, for the highest redshift galaxies
($z$ $>$ 0.9), we used the method defined by P\'erez-Montero et
al. (2007) and based on the [\oii] $\lambda$ 3727  and [\neiii] $\lambda$ 3869
emission lines in relation to the brightest available Balmer hydrogen
emission line. This diagnostic also utilizes the empirical relation between
the emission line fluxes of [\oiii] $\lambda$ 5007 and the [\neiii] line at a bluer
wavelength for a sample of well-characterised ionised gaseous nebulae of the
Local Volume. The ratio found by P\'erez-Montero et al. (2007)
[F([\oiii])/F([\neiii]) $\approx$ 15.37] agrees within the errors
with the ratio measured in those 20k zCOSMOS galaxies with
a trustable measurement of the two lines. For those galaxies with no
measurement of \hb, we used \hg\ or \hd\ and the theoretical
relations between Balmer lines from Storey \& Hummer (1995)
coefficients. We also used this method for those objects in
the range 0.5 $<$ $z$ $<$ 0.9 with no confident measurement
of the [\oiii] emission lines.
Regarding extinction correction, unlike the previous redshift range
where the ratio of equivalent widths was used, here this does not work
due to the stellar continuum variations at [\neiii] wavelength,
according to P\'erez-Montero et al. (2007). Instead, we used the Balmer decrement
in those galaxies with more than one trustable hydrogen emission-line and,
otherwise, we took the derived inner stellar extinction from the spectral
synthesis fitting.
We also took the separation curve proposed by
Marocco et al. (2011), as shown in the lower right-hand panel of
Fig. \ref{diagn}. The number of rejected objects in this regime is
much higher than in the other three (190, or 46.0\% of the sample),
but this cannot be taken as a much higher fraction of AGNs in this
redshift range. On the contrary, it is probably a selection effect
as AGNs have higher S/N for the involved lines than pure
star-forming galaxies. In fact, increasing the S/N cut
to 3, 5, and 10  increases the fraction of AGNs to 53\%,
64\%, and 80\%, respectively.

%-----------------------------------------------------------
\begin{table}
\caption{Relative number of star-forming galaxies and narrow-line AGNs in
each redshift bin in the zCOSMOS bright sample.}
\begin{center}
\begin{tabular}{lcccc}
\hline
\hline
$z$ bin & Total & SF & NL-AGN & (\%) \\
\hline
0.0 - 0.2 &  485  & 460  & 25 & 5.2 \\
0.2 - 0.4 &  1637 & 1487 & 150 & 9.2  \\
0.4 - 0.6 &  766  & 640 & 126 & 16.5 \\
0.6 - 0.8 & 1799 & 1590 & 209 & 11.6  \\
0.8 - 1.0 & 568 & 445 & 123 & 21.7 \\
1.0 - 1.32 &  76 & 46 & 30  &  39.5 \\
\hline
All & 5331 & 4668 & 663 & 12.4 \\
\end{tabular}
\end{center}
\label{tablelim}
\end{table}
%-----------------------------------------------------------------------
\subsection{Stellar mass}

%______________________________________________ 
   \begin{figure}[h]
%   \centering
 \includegraphics[width=12cm,clip=]{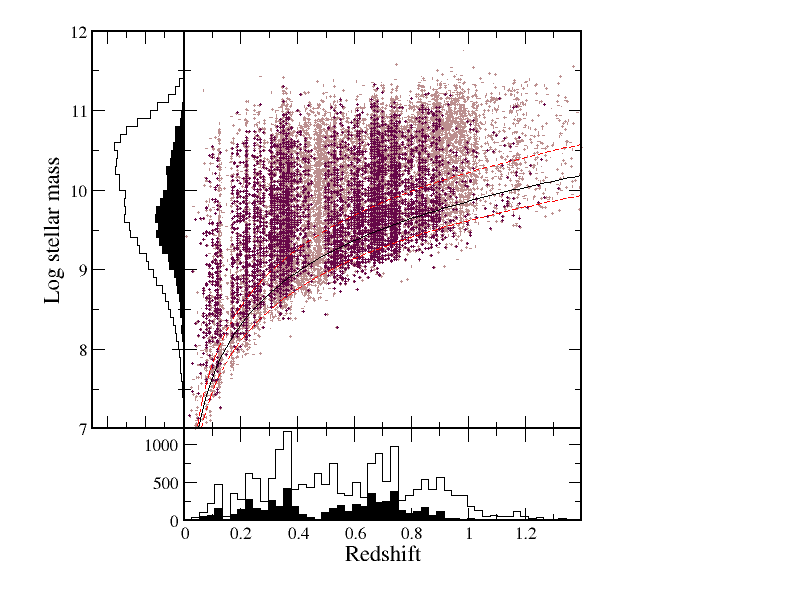}
     \caption{Relation between redshift and stellar mass (in solar masses) for the
     complete zCOSMOS 20K sample (brown points) and star-forming selected
     galaxies (violet points). The solid lines show the logarithmic fitting to the
     limiting masses of the star-forming sample for levels 25\%, 50\%, and 75\% of
     completeness. Lower and left hand panels also show the distributions for
     both the complete sample (empty histogram) and the star-forming galaxies (filled histogram)
     of $z$ and M$^*$, respectively.}
	\label{limmass}
    \end{figure}
%____________________________________

Total stellar masses for galaxies in the SDSS parent sample were taken from the
MPA/JHU catalogue, based on fits to the photometric SED.
Regarding masses in the zCOSMOS sample, they were derived from fitting stellar population synthesis
models to both the broad-band optical/near-infrared (CFHT:
u, i, Ks; Subaru: B, V, g, r, i, z; Capak et al. 2007) and far-infrared
(Spitzer/IRAC: 3.6$\mu$m, 4.5$\mu$m; Sanders et al. 2007) photometry,
and two spectral features (\hd\ absorption line and D(4000) break)
when observed in the VIMOS spectra, using a chi-square minimisation
for each galaxy. The different methods used to compute
stellar masses, based on different assumptions about the
population synthesis models and the star formation histories, are
described in Bolzonella et al. (2010). The accuracy of the photometric
stellar masses is satisfactory overall, with typical dispersions
due to statistical uncertainties and degeneracies of the
order of 0.2 dex. The addition of secondary bursts to a continuous
star formation history produces systematically higher (up to
40\%) stellar masses, while population synthesis models taking
the TP-AGB stellar phase into account (Maraston 2005) 
systematicaly lower M$^*$ by  0.10 dex. Finally,
the uncertainty on the absolute value of M$^*$ due to assumptions
on the Initial Mass Function (IMF) is within a factor of 2 for the
typical IMFs usually adopted.
In this paper, we have adopted the stellar masses calculated
with the stellar population models of Bruzual \& Charlot (2003),
with the addition of secondary bursts to the standard declining
exponantial star formation history. 

The left hand panel of Fig. 2 shows the stellar mass distribution (in units of solar mass)
of the complete zCOSMOS 20k sample of 18\,206 galaxies with secure redshifts.
The median value of this distribution is log(M*) = 9.96. 
The stellar mass distribution
for the 4\,668 star-forming selected galaxies has a  median value
of 9.66. The lower panel in the same figure shows the redshift distributions
for the same samples, with mean values of 0.59 for the complete sample
and 0.52 for the selected star-forming galaxies.

For the analysis of stellar masses, we must consider that the minimum ``detected" stellar mass is a function of redshift. We thus defined a minimum mass for each redshift bin, which is the 
lowest mass at which the mass function can be considered reliable and unaffected by
incompleteness on mass-to-light (Ilbert et al. 2004, Pozzetti et al. 2007).
To calculate the minimum mass, we first computed the limiting mass, which is
the stellar mass that an object would have at the limiting magnitude, following the
expression
\begin{equation}
\log(M^*_{lim}) = \log(M^*) + 0.4 \cdot (I_{obs} - I_{sel})
\end{equation}
where $M^*_{lim}$ is the limiting mass in solar masses, $M^*$ the stellar mass in solar masses, $I_{obs}$ the observed
$I$-band magnitude, and $I_{sel}$ the $I$-band magnitude limit for each field, which is equal 
to 22.5 for the studied zCOSMOS sample. In Fig. \ref{limmass} we show the stellar masses of 
the selected star-forming galaxies as a function of redshift 
in the range $0 < z < 1.4$. 
In the same plot we show the logarithmic fits to the 25, 50, and 75 percentile level of the distribution
of limiting masses. These fits can be taken as estimates of the minimum masses.
In Table \ref{tablelim} we list the values of these fits for different redshift bins and they
will be considered in the following as miminum masses for different levels of completeness.

%-----------------------------------------------------------
\begin{table}
\caption{Derived minimum masses (in log(M/M$_{\odot}$)) for different redshift bins obtained
for different levels of completeness}
\begin{center}
\begin{tabular}{lccc}
\hline
\hline
$z$ bin & M$^*_{min}$ (25\%) & M$^*_{min}$ (50\%) & M$^*_{min}$ (75\%) \\
\hline
0.0 - 0.2 &  7.45 & 7.63 & 7.80 \\
0.2 - 0.4 &  8.49 & 8.70  & 8.96 \\
0.4 - 0.6 & 8.99 & 9.17 & 9.49 \\
0.6 - 0.8 & 9.28 & 9.51 & 9.85  \\
0.8 - 1.0 & 9.52 & 9.76 & 10.11 \\
1.0 - 1.32 &  9.71 & 9.95 & 10.32 \\
\hline
\end{tabular}
\end{center}
\label{tablelim}
\end{table}
%-----------------------------------------------------------------------

\subsection{Star formation rates}

%______________________________________________ 
   \begin{figure*}
   \begin{minipage}{180mm}
  \centerline{
   \includegraphics[width=10cm,clip=]{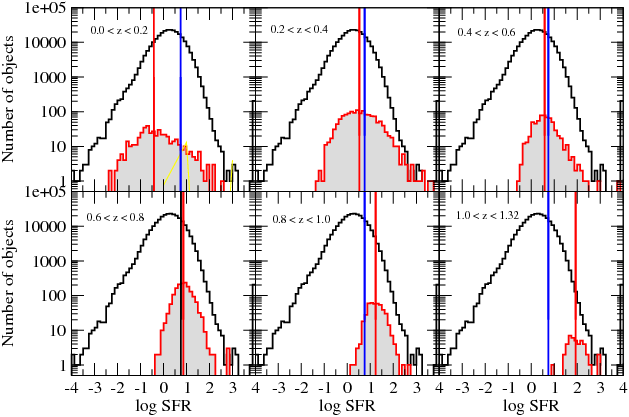}}
 	 \caption{Distribution of the calculated SFR for the star-forming
 	 galaxies of the SDSS (in black) and for zCOSMOS star-forming galaxies in different redshift bins (in red).  
	 The vertical lines show the medians for the whole zCOSMOS sample (in blue) and
 	 for each redshift range (in red).}
	 \label{sfr_hist}
 	\end{minipage}
    \end{figure*}
%____________________________________

For the parent SDSS sample, star formation rates were taken from the MPA/JHU 
catalogue. These SFRs were derived using the technique described in
Brichmann et al. (2004).
For the selected star-forming galaxies of the zCOSMOS sample, 
SFRs were calculated using the luminosity of
the brightest available Balmer emission line. 
These values were corrected for aperture effects using the factors 
derived from HST photometry. In those cases in which this value was not available, 
we used Subaru photometry. 
Extinction correction was carried out using the Balmer decrement for
those objects with more than one Balmer hydrogen recombination line with
good S/N, and assuming the theoretical ratios at standard conditions
of temperature and density from Storey \& Hummer (1995) and the
Cardelli et al. (1989) extinction law. For objects with only one
available Balmer line (a 36\% of the selected sample), we assumed a reddening 
coefficient from the E(B-V) parameter derived from the stellar synthesis fitting. The discrepancies found in those
objects with both sources of information about the inner extincion do
not vary the SFR distribution significantly.  However, it must be stressed that
extinction of the gas and stars do not have to correlate, as is the
case in this sample, so we preferred to use gas extinction whenever possible.

We derived a linear relation between H$\alpha$ luminosity and the
log(SFR) calculated by Brinchmann et al. (2004)
for the DR7-SDSS sample in the Local Volume.  This relation yields
\begin{equation}
\log ({\rm SFR}) = 0.982 \cdot \log {\rm L(H}\alpha) - 40.66
\end{equation}
and then we derived SFRs in the star-forming selected zCOSMOS galaxies.
In those redshift ranges
({\em i.e.} for $z$ $>$ 0.44) for which \ha\ is not observed in the
VIMOS spectra, we used \hb, \hg, or \hd\ emission lines instead, based on the theoretical coefficients
between Balmer hydrogen lines from Storey \& Hummer (1995) for
typical values of both electron temperature and density.

In Fig. \ref{sfr_hist} we compare the SFR distribution of the
SDSS galaxies with the  corresponding SFR 
distributions for different redshift bins in the zCOSMOS selected sample.
The vertical blue solid line shows the median SFR for the whole zCOSMOS
sample (0.75 M$_{\odot}$/yr), 
while the red vertical solid line shows 
the median SFR for each plotted bin.
As can be observed, on one hand the median value for the whole zCOSMOS 
sample is higher than for the SDSS (with a median of 0.25 M$_{\odot}$/yr) and
the median SFR in each bin increases with $z$ from
a value log(SFR) = -0.41 M$_{\odot}$/yr in the range 0 $<z<$ 0.2 up to 
log(SFR) = 1.93 M$_{\odot}$/yr for $z>$ 1.0. This increase in SFR with cosmic time 
has already been established well by several authors ({\em e.g.}
Daddi et al., 2007; Bouch\'e et al., 2010), but may be also a
selection effect when observing the brightest galaxies at larger $z$.

\subsection{Metallicity}
\label{meta}

%______________________________________________ 
   \begin{figure*}
   \begin{minipage}{180mm}
  \centerline{
   \includegraphics[width=10cm,clip=]{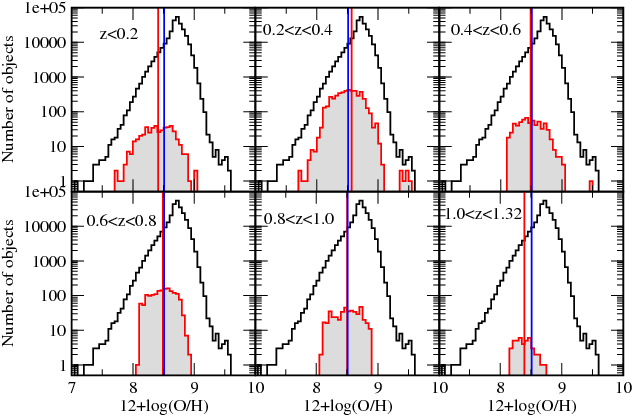}}
 	\caption{Distribution of the calculated 12+log(O/H) for the star-forming
 	 galaxies of the SDSS (black histogram) and zCOSMOS (red histograms) for different redshift bins. 
	  The vertical lines show the medians for the whole zCOSMOS sample (in blue) and
 	 for each redshift range (in red).}
	\label{Z_hist}
 	\end{minipage}
    \end{figure*}
%____________________________________

The metallicity of the studied star-forming galaxies can be estimated using
the oxygen abundances of their gas phase as a proxy. The most accurate method of
deriving this chemical abundance is based on
the previous determination of the electron temperature of the gas, via the intensity ratio
of nebular-to-auroral emission lines ({\em e.g.} [\oiii] $\lambda\lambda$ 5007, 4363) and the 
relative intensity of the strongest nebular emission lines to a hydrogen recombination line
(the so-called $T_e$-method).
However, as no information is available in the VIMOS spectra of the zCOSMOS 20k sample
about the weak auroral lines, it is necessary to apply strong-line methods. These are based on 
direct calibration of the relative intensity of the strongest collisionally excited emission lines with grids of 
photoionisation models or samples of objects with an accurate determination of
the oxygen abundance.

In this work, our aim is to derive metallicities that agree with those
derived using the $T_e$-method. Therefore, for the SDSS sample and the
zCOSMOS galaxies in the redshift range 0.01 $<$ $z$ $<$ 0.44, we used the
calibration of the N2 parameter proposed by PMC09, which is based on a sample of objects
with a good determination of 12+log(O/H) via the $T_e$-method.
The N2 parameter, defined as the ratio of [\nii] $\lambda$ 6584 to \ha\ by
Storchi-Bergmann et al. (1999), has been already used by Denicol\'o et al. (2002) as
a metallicity proxy. The main advantages of this parameter are i) its independence on
reddening and flux calibration and ii) its monotonical relation with Z over a very wide
range of metallicity.  The PMC09 calibration, as
discussed in Queyrel et al. (2012), is also consistent with other calibrations of the N2 parameter
based on photoionisation models ({\em e.g.} Denicol\'o et al., 2002;
Pettini \& Pagel, 2004; Nagao et al., 2006). According to P\'erez-Montero \& D\'\i az (2005), 
the main drawbacks of this parameter are its dependence on the ionisation parameter, the equivalent 
effective temperature, and the N/O. Taking all these effects into account, the N2 
parameter presents an overall uncertainty  of about 0.3 dex across the entire metallicity range.
This makes this parameter to be very uncertain for the determination of chemical
abundances in single objects, but it offers many advantages when it is used with
statistical purposes.

For higher redshifts, as the [\nii] emission-line is not accessible anymore in VIMOS spectra, we
derived metallicities using the \rdostres\ parameter. This was defined by
Pagel et al. (1979) as the relative sum of [\oii] $\lambda$ 3727 and [\oiii]
$\lambda\lambda$ 4959, 5007 to \hb\ intensities. Its relation with Z is bivaluated
({\em i.e.} it increases with increasing Z at low Z and it decreases
with increasing Z at high Z) and it has a strong additional dependence
on ionisation parameter and equivalent effective temperature. To minimise this dependence
we used the calibration proposed by Kobulnicky et al. (2003) based on the
photoionisation models from McGaugh (1991). This calibration takes into account the
dependence of \rdostres\ on the ionisation parameter with additional terms
as a function of the [\oii]/[\oiii] ratio. The choice of the calibration branch (low- or
high-Z) for each object was done according to the value providing less dispersion
in the resulting mass-metallicity relation.  Using this criterion, we selected
the upper-branch calibration in all cases translating into oxygen abundances with
a lower limit of 12+log(O/H) = 8.0. However, it is expected that a subsample of the objects have lower metallicities,
as those identified by Maier et al. (in prep.) using ISAAC-VLT near-IR observations 
of zCOSMOS galaxies of the bright sample  and measuring emission-lines which 
allow the choice of the appropriate branch of the \rdostres\ calibration. 
An estimate of the relative weight of these metal-poor objects in the $z$ 
bins where double-valued strong-line methods were used can be obtained from
the local SDSS sample. The proportion of objects with 12+log(O/H) $<$ 8.0 from
the N2 method is only 0.2\%, and if we take into account the minimum mass
at 50\% of completeness for $z$ $>$ 0.5, is only a 0.016\%. 

Regarding the reddening dependence for the [\oii]/\hb\ ratio, we used the same procedure based on 
equivalent widths and the factor $\alpha$ as a function of D(4000), as described in section \ref{agnsel}.

Finally, for galaxies with a redshift $z$ $>$ 0.9, we used the method described by
P\'erez-Montero et al. (2007) based on the relation between
[\neiii] $\lambda$ 3869 and [\oiii] $\lambda$ 5007 (the so-called O$_{2Ne3}$ parameter)
and the theoretical relation between \hb\ and other Balmer lines at bluer wavelengths
with enough S/N ({\em i.e.} \hg\ and \hd).
For these objects, we also considered  reddening correction from the
Balmer decrement when more than a hydrogen emission-line is available and,
otherwise, we took the stellar extinction derived from spectral synthesis fitting
as described in section \ref{agnsel}.

As already described by Kewley \& Ellison (2008), important differences can
arise between the metallicities derived using different strong-line methods and/or
different calibrations of the strong emission-line ratios. In our case, as the
PMC09 calibration of the N2 parameter is based on the compilation of objects
with a ``direct" determination of the oxygen abundance, 
while the calibration of \rdostres\ is based on sequences of photoionisation 
models, this difference can be very large.
To convert the metallicities derived from \rdostres\ and O$_{2Ne3}$ parameters to those estimated 
in the local universe from the N2 parameter, we used the following linear
relations which is based on models by Charlot \& Longhetti (2001) as described in
Lamareille (2007).
\begin{equation}
\log(\rm{O/H})_{N2} = 1.228 \times \log(\rm{O/H})_{R23} - 1.921
\end{equation}
\noindent for the lower branch calibration of \rdostres\, and
\begin{equation}
\log(\rm{O/H})_{N2} = 1.203 \times \log(\rm{O/H})_{R23} - 1.932
\end{equation}
\noindent for the calibration of the upper branch.

In Fig. \ref{Z_hist} we show the oxygen abundance distribution for the SDSS
selected star-forming galaxies compared with the same distributions for different $z$ bins
in the zCOSMOS sample in order to investigate their completeness in metallicity.  
In the lowest redshift bins ($z$ $<$ 0.4), the high-metallicity values
measured in SDSS galaxies are not reached in zCOSMOS basically because almost no 
massive galaxies are detected, as the sampled volume probed by zCOSMOS at these low 
redshifts is much smaller than for SDSS. For low metallicities, no values lower than 
12+log(O/H) $<$ 7.7 are found in zCOSMOS in the same low $z$ bins because the
minimum [\nii] flux which can be measured in this survey is brighter than in SDSS.
Nonetheless, the statistical weight of this low-metallicity queue in the SDSS sample
is negligible (0.02\%). We can thus consider that our selection of zCOSMOS galaxies 
is representative, in terms of metallicity, of the low-redshift ($z < 0.4$) star-forming galaxies.

At larger redshift ($z> 0.4$), there is also a selection effect related to the minimum emission-line flux
detected in zCOSMOS, but this time it affects the high-metallicity regime. 
Indeed, in this regime, all oxygen abundances were derived from the upper-branch calibration 
of \rdostres\ (or O$_{2Ne3}$ for $z$ $>$ 0.9), where weaker relative emission lines imply 
higher Z. Therefore, the minimum detectable [\oii] and [\oiii] emission lines 
imposes an upper limit on the metallicity of zCOSMOS galaxies. However, as in the case of
low-metallicity galaxies at low redshift, this has a negligible impact on the final distributions as 
SDSS galaxies with 12+log(O/H) $>$ 9 only represents 1.14 \%\ of the total sample. 
Regarding the cutoff seen for lower metallicity at this redshift regime, it
corresponds to the choice of the upper branch calibration in all cases
(12+log(O/H) $>$ 8.0). Although a number of galaxies is expected to have low
metallicities (as those observed with ISAAC in Maier et al. (in prep.)), these galaxies have 
a negligible statistical weight when we compare with the SDSS sample
(0.2\%). 
In any case, this number cannot be taken as an estimate of the
relative number of very-low metallicity galaxies in all the zCOSMOS sample.

\subsection{Nitrogen-to-oxygen abundance ratio}

%______________________________________________ 
   \begin{figure*}
   \begin{minipage}{180mm}
  \centerline{
   \includegraphics[width=10cm,clip=]{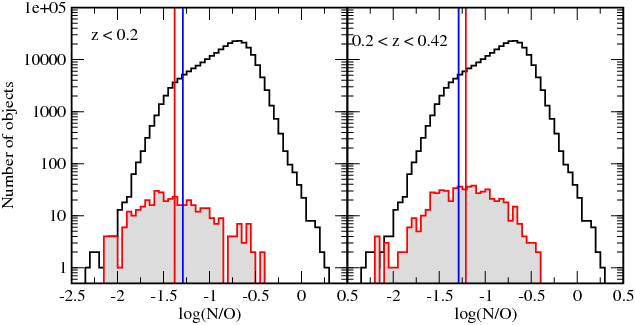}}
 	 \caption{Distribution of the calculated log(N/O) for the star-forming
 	 galaxies of the SDSS (black histogram) and zCOSMOS (red histograms) for different redshift bins. 
	  The vertical lines show the medians for the whole zCOSMOS sample (in blue) and
 	 for each redshift range (in red).}
	\label{NO_hist} 	
	\end{minipage}
    \end{figure*}
%____________________________________

The N/O was derived for both the SDSS sample
and the zCOSMOS 20k sample (up to $z\sim  0.42$ only), using the
empirical calibration of the N2S2 parameter.
This parameter is defined as the ratio of the emission-line fluxes of
[\nii] $\lambda$ 6584 and [\sii] $\lambda\lambda$ 6717, 6731, which according
to PMC09 has a linear relation with log(N/O).  Although the ratio between
[\nii] $\lambda$ 6584 and [\oii] $\lambda$ 3727, the so-called N2O2 parameter,
is also a good indicator of N/O (P\'erez-Montero \& D\'\i az, 2005; PMC09),
we did not use it for sake of consistency with the zCOSMOS data, for which
no [\oii] information exists for $z$ $<$ 0.5.

In Fig. \ref{NO_hist} we show the distribution of log(N/O) for both the SDSS and zCOSMOS samples. 
Two redshift bins are considered for the zCOSMOS sample: $z \leq 0.2$ and $0.2 < z \leq 0.42$.
Contrary to metallicity, the N/O is well sampled for low values in zCOSMOS using
the N2S2 parameter. However, this is not the case for high values of N/O. As in the
case of metallicity, this is because only few massive galaxies are sampled at
these redshifts within zCOSMOS, and thus very high values of N/O are missing in this sample.
This effect is also seen in the zCOSMOS sample because the
median log(N/O) for the lower redshift bin (-1.38) is lower than for $z$ larger than 0.2
(-1.29), and both of them are much lower than the median log(N/O) for
the entire SDSS sample (-0.83) since more massive galaxies are sampled on it.

\section{Results and discussion}

\subsection{Cosmic evolution of the mass-metallicity relation}

%______________________________________________ 
   \begin{figure}[h]
%   \centering
 \includegraphics[width=9cm,clip=]{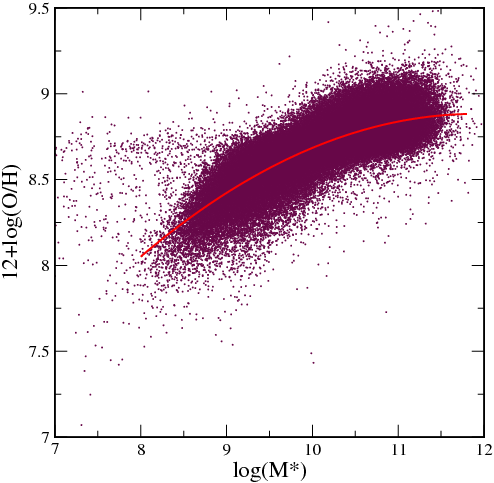}

      \caption{Relation between stellar mass (in units of M$_{\odot}$) and 12+log(O/H) for
      the selected star-forming galaxies of the SDSS-DR7. The red solid line shows the quadratical
      fit to the medians in bins of 0.2 dex of stellar mass.}
	\label{sdss_mz}
    \end{figure}
%____________________________________

To consistently study the cosmic evolution of the $MZR$ 
we derived its shape in the selected star-forming galaxies of the 
SDSS, which are representative of the local universe. As described in the sections above 
we used the compiled stellar masses from the MPA/JHU catalogue and
calculated oxygen abundances from the PMC09 calibration of the N2 parameter.  
The derived $MZR$ for the SDSS is shown in
Fig. \ref{sdss_mz}. We computed a quadratical
fit to the median values of O/H per stellar mass bins of 0.2 dex. The resulting fit has
the following expression
\begin{equation}
y = 1.1598 + 1.2971\cdot x - 0.0545 \cdot x^2
\end{equation}
\noindent where $y$ is 12+log(O/H) and $x$ is 
log(M$_*$) in units of M$_{\odot}$. This fit is valid in the range
of stellar mass 8.0 $<$ log(M/M$_{\odot}$) $<$ 11.8.
The dispersion was calculated as the
standard deviation of the residuals to the resulting fit, giving a result
of 0.090 dex. This dispersion is slightly larger for log(M$_*$) $<$ 10
(0.105 dex) than for higher stellar masses (0.081 dex). We checked that
our quadratical fit is consistent with the results given by Kewley \& Ellison (2008)
for a similar subsample of the SDSS. By converting our oxygen abundances
to those studied in that work using different strong-line methods or calibrations of
N2, we measured that our median Z is 0.10 dex higher on average for all stellar
masses. However, we also found that by increasing the S/N cutoff of our sample
to 10, the agreement with the fittings of the $MZR$ provided by Kewley \& Ellison (2008)
is perfect. Therefore, we conclude that using a high S/N cutoff introduces a selection effect 
in the subsample, which mainly affects to the metal-rich galaxies when it is applied
to all the involved emission lines.

%-----------------------------------------------------------
\begin{table*}
\begin{minipage}{180mm}
\caption{Average offsets between the medians
measured in the SDSS reference sample and for the star-forming galaxies of zCOSMOS 20k in each
redshift bin for the mass-metallicity relation ($MZR$), the star formation-corrected mass-metallicity
relation ($SMZ$), and the relation between stellas mass and N/O ($MNOR$). 
The dispersions of the fits are also given.}
\begin{center}
\begin{tabular}{lccccccc}
\hline
\hline
$z$ bin & $\Delta$ $MZR$ & $\sigma$ $MZR$ & $\Delta$ $MZR$\footnote{with Balmer lines with S/N $>$ 5} &  $\Delta$ S$MZR$ & $\sigma$ $SMZ$ & $\Delta$ $MNOR$ & $\sigma$ $MNOR$ \\
\hline
SDSS      & 0.00 & 0.09 & -- & 0.00 & 0.08 & 0.00 & 0.14 \\
0.0 - 0.2 &  +0.01 & 0.16 & +0.01 & +0.01 & 0.15 & 0.00 & 0.23 \\
0.2 - 0.4 &  -0.04 & 0.20  & -0.02 & -0.01 & 0.19 & -0.14\footnote{up to $z$ $\approx$ 0.42} & 0.31$^a$ \\
0.4 - 0.6 & -0.14  & 0.19 & -0.11 & -0.12 & 0.19 & -- & -- \\
0.6 - 0.8 & -0.19 & 0.17 & -0.12 & -0.13 & 0.17 & -- & --  \\
0.8 - 1.0 & -0.17 & 0.19 & -0.08 & -0.10 & 0.19 & -- & -- \\
1.0 - 1.32 &  -0.36 & 0.15 & -0.27 & -0.30 & 0.15 & -- & -- \\
\hline
\end{tabular}
\end{center}
\label{offsets}
\end{minipage}
\end{table*}

%______________________________________________ 
   \begin{figure*}
   	\begin{minipage}{180mm}
  \centerline{
   \includegraphics[width=6cm,clip=]{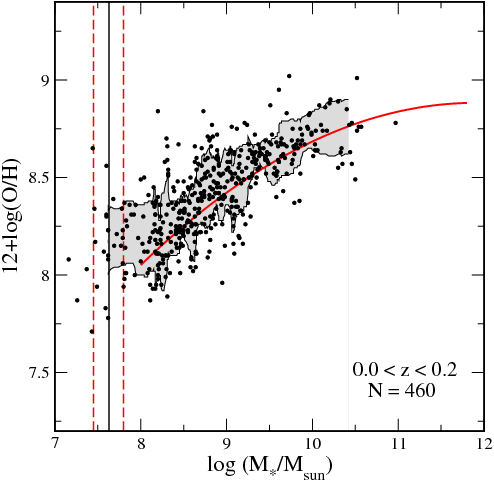}
   \includegraphics[width=6cm,clip=]{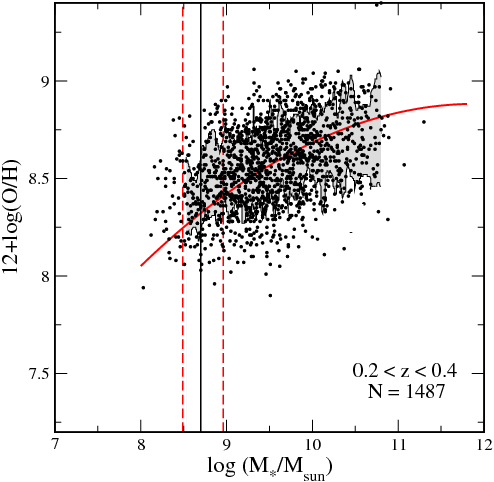}
   \includegraphics[width=6cm,clip=]{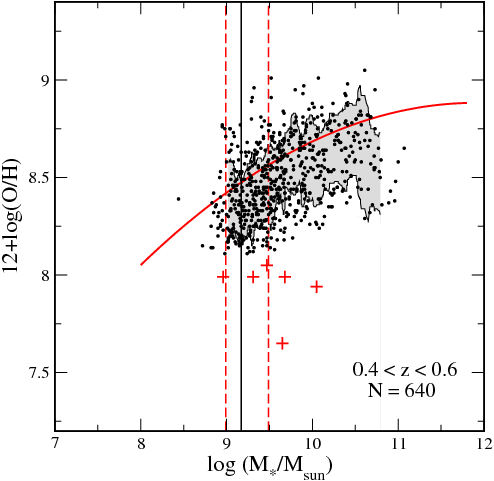}}
   \centerline{
   \includegraphics[width=6cm,clip=]{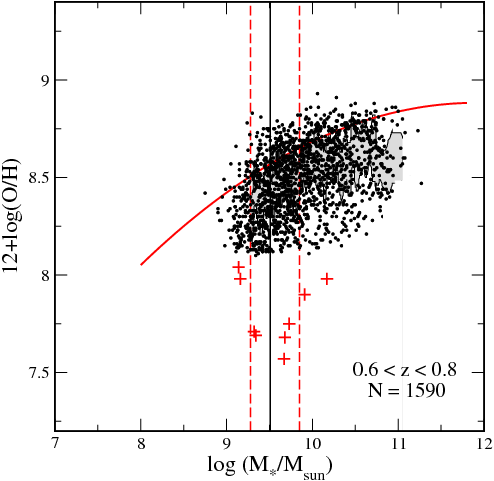}
   \includegraphics[width=6cm,clip=]{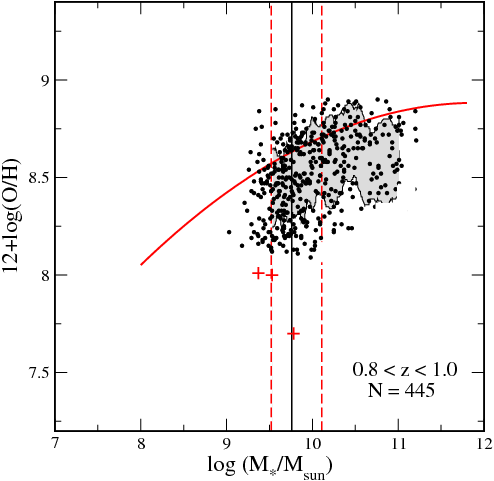}
   \includegraphics[width=6cm,clip=]{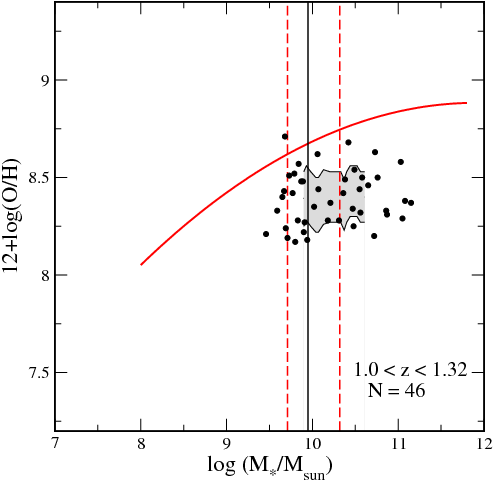}}
      \caption{Cosmic evolution of the $MZR$ for six different redshift bins increasing
      from left to right and from up to down. The red solid line shows the fit to the $MZR$ in
 the SDSS data. Grey areas show the $\pm \sigma$ intervals to the Z medians fits in different 
 mass bins for each redshift range. The vertical lines show
the minimum masses for 25\%, 50\% and 75\% levels of completeness. Red crosses 
represent low-Z objects in the sample found using ISAAC near-IR observations (Maier et
al. in prep.)}
	\label{zc_mz}
	\end{minipage}
    \end{figure*}
%____________________________________

In Fig. \ref{zc_mz}, we show the $MZR$ for six different redshift
bins between 0.01 and 1.32 and compared to the local $MZR$ derived above.
Each panel also shows the $\pm \sigma$ curves of the medians
of Z in each mass bin and the minimum masses for 
different levels of completeness as derived in section 3.2.
In the three panels correspondinbg to $z$ between 0.4 and 1.0 we show 
the small subsample of zCOSMOS galaxies observed in the near-infrared (Maier et al., in prep.) 
and for which oxygen abundances from the lower calibration of
the \rdostres\ parameter have been derived.

In Table \ref{offsets}, we summarise for each
redshift bin the vertical offset between the local $MZR$ and the average of
medians for different mass intervals in each $z$ bin. We also give the corresponding 
dispersions, calculated as the standard deviation of all the medians.
In all cases, for the sake of consistency, these offsets were calculated taking from the SDSS those
subsamples of objects in the same metallicity and stellar mass range measureable in
each $z$ bin of zCOSMOS. 
The inspection of the values listed in that table leads to the following.
(i) The derived median oxygen abundance for the same bins of stellar mass increases with
cosmological age so that it was around half the present value at $z \approx 1$ ( $\approx$ 8 Gyr ago).
(ii) The increase in the oxygen abundance with cosmic time is not uniform but,
on the contrary, it looks to be significantly larger above $z \approx$ 0.5.
(iii) Both the dispersion associated with the derived medians, and the oxygen abundance 
uncertainties intrinsic to the strong-line methods for each $z$ bin, are
larger than the offset for $z < 1.0$.

Some authors ({\em e.g.} Kobulnicky et al. 1999) have warned that taking too
low a threshold of S/N for the Balmer emission lines introduces 
significant uncertainties in  the determination of metallicity using
strong-line methods, such as \rdostres. On the contrary, as explained in previous
sections, taking too high a threshold for certain lines can introduce a 
bias in the selected sample, To quantify how changing the S/N threshold for
Balmer lines affect the median Z in the different redshift bins,
we calculated them taking only Balmer emission-lines with S/N $>$ 5, instead of 2. 
Our results indicate that the number of selected galaxies is considerably
decreased for the high $z$ bins (more than 60\% for $z$ $>$ 1 and more than 35\% for
$z$ $>$ 0.4). In Table \ref{offsets} we also list the Z offsets in relation to the SDSS $MZR$.
All of them give higher Z than in relation to the samples with S/N threshold of two for
all the involved emission lines, and this difference increases with $z$, as an increasing
number of low-Z galaxies is excluded in each bin. However, the trend to finding
lower median Z with higher $z$ is maintained, although serious caveats must be
taken for the intensity of this evolution when selection effects due
to different S/N threshold criteria are imposed.

\subsection{The SFR-corrected mass-metallicity relation and its evolution}

%______________________________________________ 
   \begin{figure*}
   \begin{minipage}{180mm}
%   \centering
 \includegraphics[width=8cm,clip=]{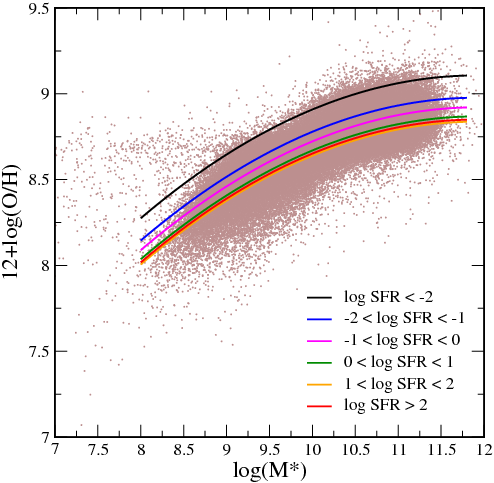}
 \includegraphics[width=8cm,clip=]{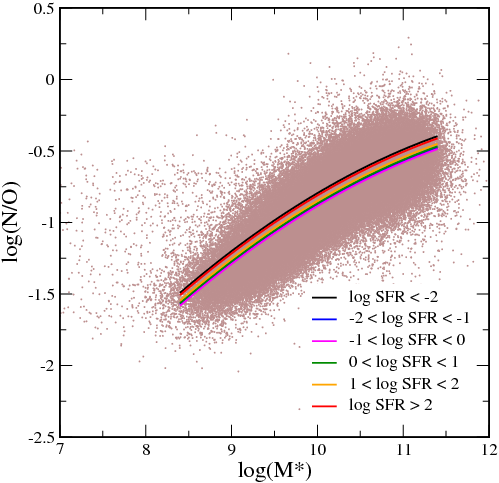}

      \caption{To the left $MZR$ and quadratical fits for different bins of 
	log SFR. To the right, same plot for the relation between log (N/O) and stellar mass.}
	\label{sdss_sfr}
	\end{minipage}
    \end{figure*}
%____________________________________

%______________________________________________ 
   \begin{figure}
%   \centering
 \includegraphics[width=9cm,clip=]{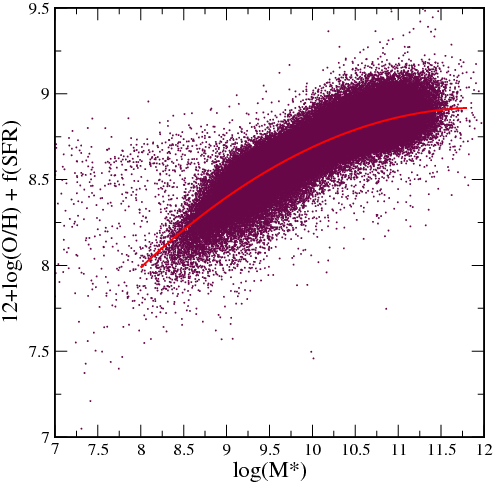}
      \caption{Relation between stellar mass in units of M$_{\odot}$ and 12+log(O/H) 
      with the correction depending on log(SFR) for
      the star-forming selected galaxies of the SDSS-DR7. The red solid line shows the quadratical
      fit to the medians in bins of 0.2 dex of stellar mass.}
	\label{sdss_smz}
    \end{figure}
%____________________________________

Stellar mass and metallicity are two of the main fundamental parameters
of galaxies. Their relationship for complete samples of galaxies, as shown in the 
above section, gives us valuable information about the main mechanisms
governing the evolution of galaxies. However, recent works
({\em e.g.} Mannucci et al., 2010; Lara-L\'opez et al. 2010) 
have demonstrated the important role of SFR, 
which is also tightly related to stellar mass and metallicity and thus affects 
the shape and the dispersion of the $MZR$.  For instance, galaxies with higher SFR have
systematically lower metallicities. This effect could be explained by invoking 
the interaction of galaxies with inflows of unenriched gas that triggers
the episodes of star formation and reduces the relative abundance of oxygen
in the gas phase. For this reason, Mannucci et al. (2010) defines a
fundamental metallicity relation ($FMR$) that is basically a surface in
the three-dimensional space formed by stellar mass, Z, and SFR in which
neither selection effects due to SFR appear nor evolution with $z$ are
detected up to $z\sim$ 2.5. This lack of evolution in the defined 3D surface (although
other authors, such as Lara-L\'opez et al. (2010) defined it as a plane) is also
found for a subsample of the zCOSMOS-bright galaxies in the redshift range
$0.2 < z < 0.8$ (Cresci et al. 2012). However, in this paper, since we try to isolate the metallicity
evolution with cosmic age independently of the stellar mass, we redefine the
question, isolating the dependence of Z with SFR to remove the
selection effects at high $z$.
In the left hand panel of Fig. \ref{sdss_sfr},  we show the $MZR$ with
different quadratical fits for different bins of SFR for the aforementioned
selected star-forming galaxies of the SDSS. 
As mentioned above, there is a
trend toward obtaining lower 12+log(O/H) for higher SFRs. 
We then applied the SFR correction only to the metallicity, in order 
to remove possible selection effects at higher $z$.  We thus used the SFR-corrected 
$MZR$ (hereafter $SMZ$ relation).

By inspecting Fig. \ref{sdss_sfr} again, we clearly see that the evolution of the vertical
offsets between the quadratical fits to subsamples with different SFRs is not linear.
Indeed, we can fit a quadratical SFR-correction to the Z axis 
in order to reduce the dispersion and remove the dependence on SFR.
The additional term to add to the metallicity axis is
\begin{equation}
f(\log \textrm{SFR}) = 0.0505\cdot (\log \textrm{SFR}) - 0.0144\cdot (\log \textrm{SFR})^2.
\end{equation}

In Fig. \ref{sdss_smz} we show the corrected $MZR$,
in which the corrected metallicity is represented as a function of the stellar mass.
The resulting quadratical fit to the Z medians in bins of 0.2 dex
in stellar mass in the range 8.0 $<$ log(M/M$_{\odot}$ $<$ 11.8 is 
\begin{equation}
y = 0.4655 + 1.4118\cdot x - 0.0589\cdot x^2
\end{equation}
where $y$ = 12+log(O/H)+$f$(SFR) and $x$ is the stellar mass in
units of solar masses.
The dispersion of this new relation is slightly lower than for the uncorrected
$MZR$ (0.083 dex) and, as in the uncorrected case, the dispersion is lower 
for masses higher than log(M$_{*}$) = 10.0 (0.075 dex) than for lower masses
(0.097dex). An explanation of the fact that the dispersion is only reduced by 
0.01 dex at all stellar masses can be found the majority of
SDSS galaxies not having extreme SFR values. The dispersion in the SFR 
distribution of the SDSS sample is only 0.37 dex. A much more thorough calculation 
of this correction should be done using well-known galaxies with extreme values 
of the SFR to put more precise constraints on the SFR-correction term of the $SMZ$.

%______________________________________________ 
   \begin{figure*}
   \begin{minipage}{180mm}
  \centerline{
   \includegraphics[width=6cm,clip=]{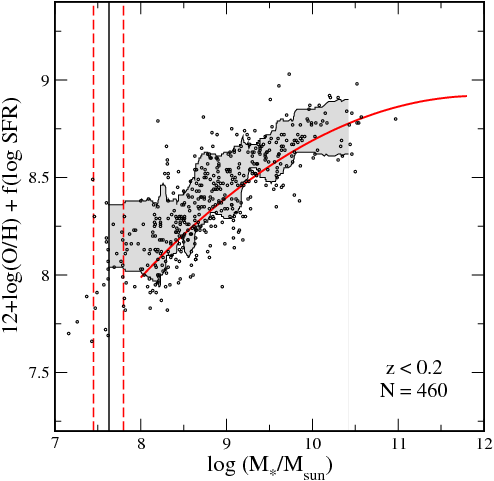}
   \includegraphics[width=6cm,clip=]{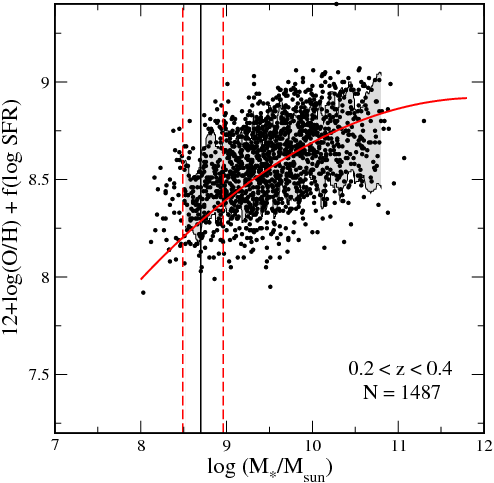}
   \includegraphics[width=6cm,clip=]{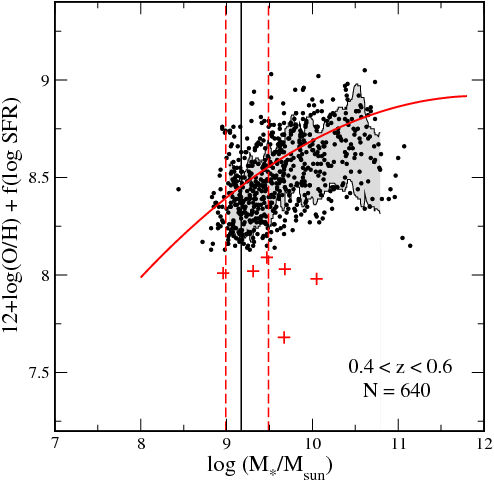}}
   \centerline{
   \includegraphics[width=6cm,clip=]{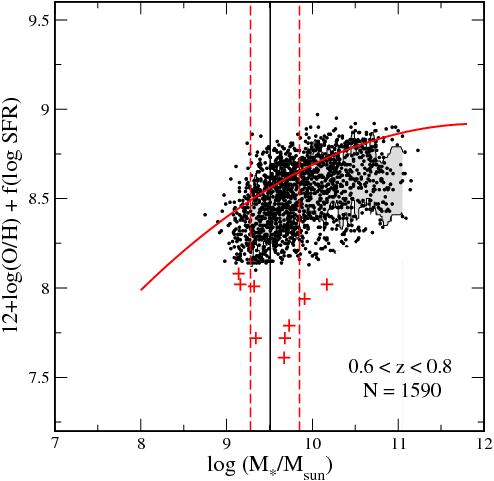}
   \includegraphics[width=6cm,clip=]{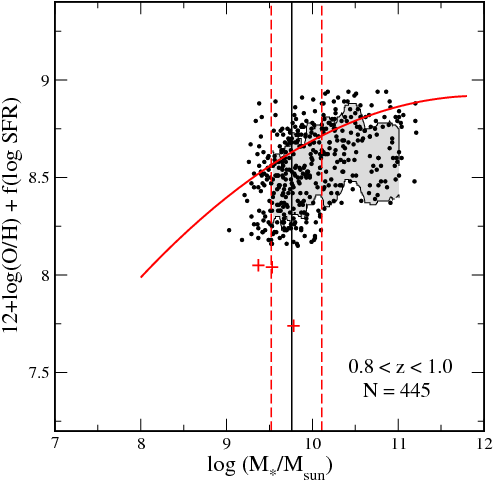}
   \includegraphics[width=6cm,clip=]{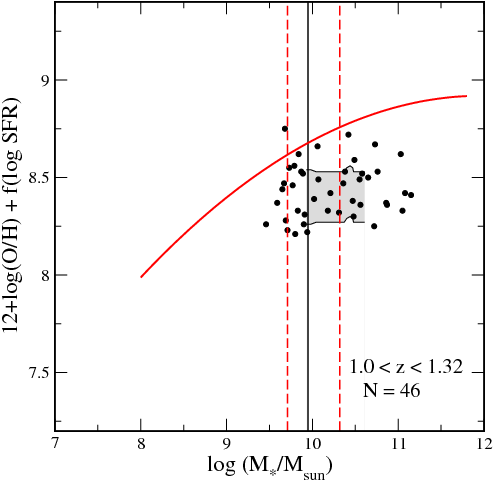}}
      \caption{Cosmic evolution of the SFR-corrected $MZR$ for six different redshift bins increasing
      from left to right and from up to down. The red solid line shows the fitting to the $SMZ$ in the
      SDSS data. Upper and lower solid lines show the $\pm \sigma$ intervals to the fitting 
to the $y$ medians in different mass bins for each redshift range. The vertical lines show
the minimum mass limits for 25\%, 50\% and 75\% levels of completeness. Red crosses 
represent low-Z objects in the sample found using ISAAC near-IR observations (Maier et
al. in prep.)}
	\label{zc_smz}
	\end{minipage}
    \end{figure*}
%________________________

To study to what extent selection effects of the SFR can affect the
conclusions reached from the study of the $MZR$, we applied the SFR-correction to
the selected zCOSMOS galaxies at different $z$ bins. 
The $SMZ$ is plotted in Fig. \ref{zc_smz} for the same six different redshift
bins between 0.01 and 1.32 with respect to the relation derived for the SDSS sample.
The panels also show the $\pm 1\sigma$ bands around the fitting to the medians
for Z+$f$(SFR) in each mass bin and the minimum mass for 
different levels of completeness. In Table \ref{offsets}, we summarise for each
redshift bin the vertical offset between the reference SDSS $SMZ$ and the average of
median offsets for each mass bin and the dispersion, calculated as the standard 
deviation of the residuals to the fitting. 

The inspection of the values listed in that table leads to the following conclusions.
(i) There is still a trend toward lower median values of the SFR-corrected Z for
higher $z$, although the measured offsets are lower. This implies that,
although the $MZR$ is affected by SFR selection effects for samples taken 
at high $z$, these are not sufficient to explain the observed evolution, as
stated in other studies ({\em e.g.} Mannucci et al. 2010; Cresci et al. 2012).
(ii) The dispersion of the SFR-corrected Z is not noticeably lower in each
$z$ bin than for the $MZR$, possibly because that the SFR distributions for each
$z$ bin do not have very large dispersions, so the correction affects the median value 
but not the dispersion of each distribution.

\subsection{The relation between stellar mass and nitrogen-to-oxygen ratio
and its evolution}

%______________________________________________ 
   \begin{figure}[h]
%   \centering
 \includegraphics[width=9cm,clip=]{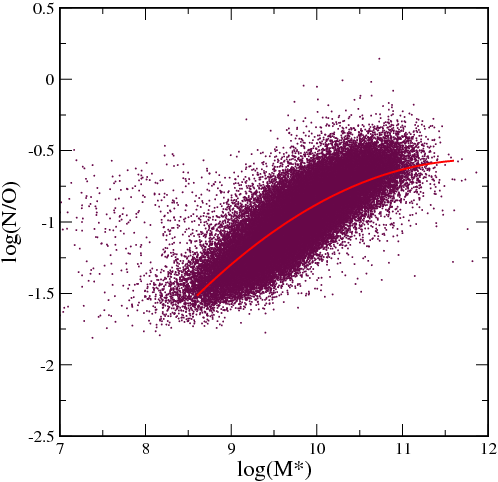}

      \caption{Relation between stellar mass in units of M$_{\odot}$ and log(N/O) 
      for the star-forming selected galaxies of the SDSS-DR7. The red solid line shows the quadratical
      fit to the medians in bins of 0.2 dex of stellar mass.}
	\label{sdss_mno}
    \end{figure}
%____________________________________

The N/O complements the study of the chemical composition of
the nebular phase of star-forming galaxies, and it is a robust indicator of the
chemical evolution of galaxies. Besides, as the production rate of secondary
N is faster than for O (Henry et al. 2000; Thuan et al. 2010), the amplitude
of the evolution of its relation with stellar mass gives 
more reliability to the conclusions reached about its evolution with cosmic age.

Although the relation between stellar mass and N/O  (hereafter $MNOR$) 
has been already studied by PMC09, we re-calculated this relation using the same selected 
star-forming galaxies of the SDSS in order to be compared with the zCOSMOS galaxies at 
higher $z$. In Fig. \ref{sdss_mno} we show this relation between stellar mass and
log(N/O) for the selected star-forming galaxies of the SDSS-DR7. As explained above, 
the N/O was calculated using the PMC09 calibration of the N2S2 parameter.
As in the case of the $MZR$, there is a trend toward finding higher N/O for more massive
galaxies because all galaxies in this sample already lie in the regime of production
of secondary N and so  its N/O increases with metallicity.
We performed a quadratical fit to the medians of
N/O in bins of stellar mass of 0.2 dex in the stellar mass range 
8.6 $<$ log(M/M$_{\odot}$) $<$ 11.6 which gives the following expresion:
\begin{equation}
y = -9.4457 + 1.3577\cdot x - 0.0501\cdot x^2
\end{equation}
\noindent where $y$ is log(N/O) and $x$ is
log(M$_*$) in units of solar masses. The dispersion, calculated as the
standard deviation of the residuals, is equal to 0.144 dex.
Contrary to the $MZR$ relation, this dispersion is slightly lower for masses
greater than 10$^{10}$ solar masses (0.136 dex) than for
lower masses (0.156 dex). The more restricted stellar mass range where the
fit is possible and the higher dispersion for lower stellar masses are possibly owing
to the presence of objects with additional primary N production at the low
stellar mass regime.

Besides, unlike the $MZR$ too, there is no apparent selection effect
due to different values of the SFR, as can be seen in right hand panel of 
Fig. \ref{sdss_sfr}. This lack of SFR-dependence of the relation between stellar mass
and N/O could be a confirmation that the relation between Z and SFR is
mainly due to the inflows of metal-poor gas, which would trigger the star formation,
decreasing the relative abundance of the metals but which, in contrast, do not
affect the ratio between secondary and primary elements (Edmunds 1990).

%______________________________________________ 
   \begin{figure*}
   \begin{minipage}{180mm}
  \centerline{
   \includegraphics[width=6cm,clip=]{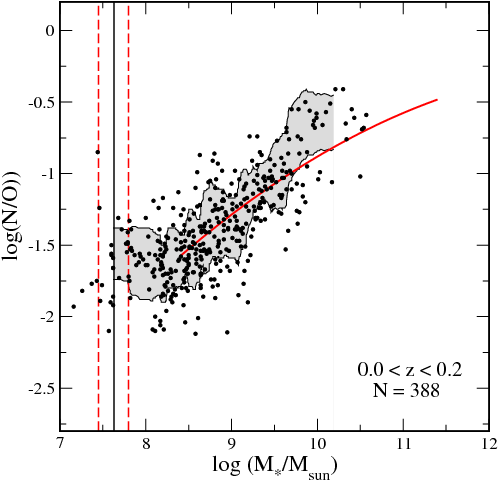}
   \includegraphics[width=6cm,clip=]{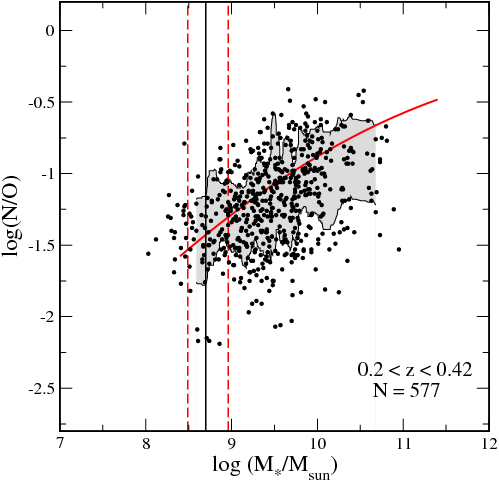}}

      \caption{Cosmic evolution of the $MNOR$ for two different redshift bins.
      The red solid line shows the fit to the $MNOR$ in the
      SDSS data. Upper and lower solid lines show the $\pm \sigma$ intervals to the fitting 
to the N/O medians in different mass bins for each redshift range. The vertical lines show
the minimum mass limits for 25\%, 50\% and 75\% levels of completeness}
	\label{zc_mno}
	\end{minipage}
    \end{figure*}
%____________________________________

%______________________________________________ 
   \begin{figure*}
   \begin{minipage}{180mm}
  \centerline{
   \includegraphics[width=6cm,clip=]{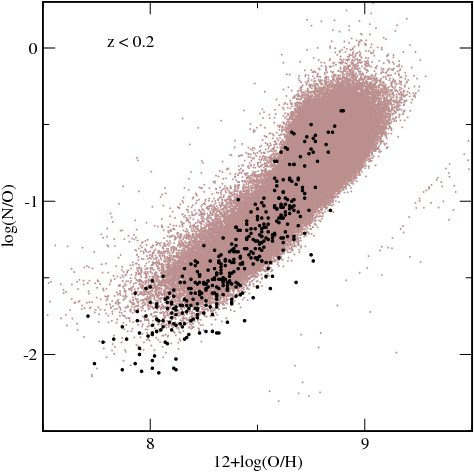}
   \includegraphics[width=6cm,clip=]{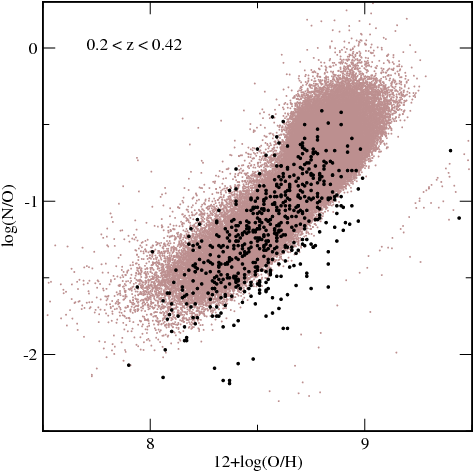}}

      \caption{Relation between 12+log(O/H) and log(N/O) for the star-forming
selected SDSS galaxies (in brown) and for zCOSMOS for $z < 0.2$ (black dots in
left hand panel) and for $0.2 < z < 0.42$ (right hand panel). }
	\label{zc_OH_NO}
	\end{minipage}
    \end{figure*}
%__________________

Therefore, the study of the evolution of the $MNOR$ with cosmic time 
represents a robust tool for studying the evolution of the chemical content of
galaxies as a function of the age of the Universe, as N/O does not
depend on selection effects related to SFR. In Fig. \ref{zc_mno}, 
we show the evolution of this relation for the two redshift bins in zCOSMOS 
for which we were able to derive the N/O using the N2S2 parameter. 
The offsets between the relation derived in the SDSS and for the two 
analysed $z$ bins and their dispersions are listed in Table \ref{offsets}. 
The analysis of the obtained values leads to the following conclusions.
(i) For a given stellar mass, there is a significant evolution in the median N/O 
abundance ratio of galaxies from $z\sim 0.4$ up to now. 
(ii) This evolution is a consequence of the evolution of the metallicity of galaxies
with cosmic age. As can be seen in Fig. \ref{zc_OH_NO}, the average relation 
between 12+log(O/H) and log(N/O) in the regime of production
of secondary nitrogen is kept in the two redshift bins.
(iii) This evolution cannot be due to any selection effect of the sampled galaxies
at high redshift that have, on average, larger SFR, because N/O is not
affected by inflows of metal-poor gas in galaxies.
(iv) In the secondary nitrogen production regime, N/O increases linearly
with oxygen abundance, as can be seen in Fig.~\ref{zc_OH_NO}. Therefore
the N/O evolution derived from $z$ $\approx$ 0.3 can be transformed into
metallicity evolution using the slope of this linear relation. This slope
is around 1.4 for both the two redshift bins analysed, so this gives a variation
in 12+log(O/H) of 0.1 dex on average, which is slightly higher than
the values found using both $MZR$ and $SMZ$. However, it must be taken
the large scatter into account in the N/O vs. O/H when computing the
O/H evolution in this way

\section{Summary and conclusions}

We studied the cosmic evolution of the $MZR$ from $z$ $\approx$ 1.3 up to now
using complete samples of star-forming galaxies: in the Local Universe  we used 
the SDSS-DR7 and for higher $z$ we used the 20k zCOSMOS bright sample.
The selection of star-forming objects and derivation of physical properties,
including stellar masses, star formation rates, oxygen abundances, and 
N/O, were performed using consistent methods.
This is especially relevant in the case of the VIMOS spectra taken for the zCOSMOS sample 
since the set of available emission lines varies in each redshift regime.
Two important aspects were considered when deriving chemical
abundances: i) all strong-line methods used to derive 12+log(O/H) and log(N/O)
were consistent with the $T_e$-method as used in PMC09, and ii) the S/N threshold 
to select objects was put at a lower value to avoid possible selection effects
in the final distributions of the derived properties.

The analysis of the evolution of the $MZR$ in different redshift bins confirms
an increase in the metallicity from $z$ = 1.32, but the dispersion is too high 
to assess any variation in the shape of the $MZR$.

Consideringthe dependence of metallicity on SFR found by several
authors, we derived an SFR-corrected $MZR$.
This correction slighly reduces the dispersion of the $MZR$, but
not noticeably since the dispersion in the SFR distributions is not very high.
We still see the evolution of the metallicity for galaxies of the same stellar
mass bin at different epochs, although this is slightly lower than
for the $MZR$.

Finally, we studied the evolution of the $MNOR$ taking the log(N/O) instead of
(O/H) in the $MZR$. Since N/O was derived from the N2S2 parameter as calibrated
by PMC09, we carried out this study for the zCOSMOS 20k sample only up to $z$ $\approx$
0.42. The main advantage to considering the $MNOR·$ with respect to the $MZR$ is 
the independence of this relation on SFR, which confirms that the dependence
of Z on SFR is probably due to inflows of metal-poor gas in some starburst events.
In the studied redshift regime we still see the evolution of the chemical content
of galaxies with the same stellar mass from younger ages of the Universe,
confirming that the aforementioned evolution of the $MZR$ is not just a
selection effect of the observed objects at high $z$.

\begin{acknowledgements}
We thank C. Tremonti for having made the original {\em platefit} code available to the zCOSMOS collaboration.\\
This work has been partially supported by the CNRS-INSU and its Programmes Nationaux de Galaxies et de Cosmologie (France),
and by projects  AYA2007-67965-C03-02 and AYA2010-21887-C04-01
of the Spanish National Plan for Astronomy and Astrophysics, and by the project TIC114  {\em Galaxias y Cosmolog\'\i a} of the
 Junta de Andaluc\'\i a (Spain).\\
The VLT-VIMOS observations were carried out on guarateed time (GTO) allocated by the European
Southern Observatory (ESO) to the VIRMOS consortium, under a contractual agreement between the Centre
National de la Recherche Scientifique of France, heading a consortium of French and Italian institutes,
and ESO, to design, manufacture, and test the VIMOS instrument.\\
The work was based on observations obtained with MegaPrime/MegaCam, a joint project of CFHT and CEA/DAPNIA, at the Canada-France-
Hawaii Telescope (CFHT) which is operated by the National Research Council (NRC) of Canada, the Institut
National des Science de l'Univers of the Centre National de la Recherche Scientifique (CNRS) of France and
the University of Hawaii. This work is based in part on data products produced at TERAPIX and the Canadian
Astronomy Data Centre as part of the Canada-France-Hawaii Telescope Legacy Survey, a collaboration
project of NRC and CNRS.
We finally thank Jos\'e M. V\'\i lchez and Ricardo Amor\'i\ n for very fruitful discussions
that have helped to reach some of the conclusions depicted in this work.
\end{acknowledgements}


\begin{thebibliography}{}

\bibitem[Amor{\'{\i}}n et al.(2010)]{2010ApJ...715L.128A} Amor{\'{\i}}n, 
R.~O., P{\'e}rez-Montero, E., 
\& V{\'{\i}}lchez, J.~M.\ 2010, \apjl, 715, L128 
\bibitem[Bertoldi et al.(2007)]{2007ApJS..172..132B} Bertoldi, F., Carilli, 
C., Aravena, M., et al.\ 2007, \apjs, 172, 132 
\bibitem[Bolzonella et 
al.(2010)]{2010A&A...524A..76B} Bolzonella, M., Kova{\v c}, K., Pozzetti, L., et al.\ 2010, \aap, 524, A76 
\bibitem[\protect\citeauthoryear{Bouch{\'e} et 
al.}{2010}]{2010ApJ...718.1001B} Bouch{\'e} N., et al., 2010, ApJ, 718, 
1001 
\bibitem{} Brinchmann, J., Charlot, S., White, S.~D.~M., et al. 2004, MNRAS, 351, 1151
\bibitem[Brisbin 
\& Harwit(2012)]{2012ApJ...750..142B} Brisbin, D., \& Harwit, M.\ 2012, \apj, 750, 142 
\bibitem{} Brodie, J. P. \& Huchra, J. P., 1991, ApJ, 379, 157.
\bibitem[Brooks et al.(2007)]{2007ApJ...655L..17B} Brooks, A.~M., 
Governato, F., Booth, C.~M., et al.\ 2007, \apjl, 655, L17 
\bibitem{} Bruzual, G., \& Charlot, S. 2003, MNRAS, 344, 1000
\bibitem[Capak et al.(2007)]{2007ApJS..172...99C} Capak, P., Aussel, H., 
Ajiki, M., et al.\ 2007, \apjs, 172, 99 
\bibitem[Cardelli et al.(1989)]{1989ApJ...345..245C} Cardelli, J.~A., 
Clayton, G.~C., \& Mathis, J.~S.\ 1989, \apj, 345, 245 
\bibitem{} Charlot, S., \& Longhetti, M., 2001, MNRAS, 323, 887
\bibitem[Contini et al.(2002)]{2002MNRAS.330...75C} Contini, T., Treyer, 
M.~A., Sullivan, M., \& Ellis, R.~S.\ 2002, \mnras, 330, 75 
\bibitem[Cowie \& Barger(2008)]{2008ApJ...686...72C} Cowie, L.~L., \& Barger, A.~J.\ 2008, \apj, 686, 72 
\bibitem[Cresci et al.(2012)]{2012MNRAS.421..262C} Cresci, G., Mannucci, 
F., Sommariva, V., et al.\ 2012, \mnras, 421, 262 
\bibitem[Daddi et al.(2007)]{2007ApJ...670..156D} Daddi, E., Dickinson, M., 
Morrison, G., et al.\ 2007, \apj, 670, 156 
\bibitem[Dav{\'e} et al.(2012)]{2012MNRAS.421...98D} Dav{\'e}, R., 
Finlator, K., \& Oppenheimer, B.~D.\ 2012, \mnras, 421, 98 
\bibitem[Dav{\'e} et al.(2007)]{2007EAS....24..183D} Dav{\'e}, R., 
Finlator, K., \& Oppenheimer, B.~D.\ 2007, EAS Publications Series, 24, 183 
\bibitem[De Lucia et al.(2004)]{2004MNRAS.349.1101D} De Lucia, G., 
Kauffmann, G., \& White, S.~D.~M.\ 2004, \mnras, 349, 1101 
\bibitem[de Rossi et al.(2007)]{2007MNRAS.374..323D} de Rossi, M.~E., 
Tissera, P.~B., \& Scannapieco, C.\ 2007, \mnras, 374, 323 
\bibitem[Denicol{\'o} et al.(2002)]{2002MNRAS.330...69D} Denicol{\'o}, G., 
Terlevich, R., \& Terlevich, E.\ 2002, \mnras, 330, 69 
\bibitem[Edmunds(1990)]{1990MNRAS.246..678E} Edmunds, M.~G.\ 1990, \mnras, 
246, 678 
\bibitem[Edmunds 
\& Pagel(1978)]{1978MNRAS.185P..77E} Edmunds, M.~G., \& Pagel, B.~E.~J.\ 1978, MNRAS, 185, 77P 
\bibitem[Elbaz et 
al.(2007)]{2007A&A...468...33E} Elbaz, D., Daddi, E., Le Borgne, D., et al.\ 2007, \aap, 468, 33 
\bibitem{} Erb, D.~K., Shapley, A.~E., Pettini, M., et al., 2006, ApJ, 644, 813
\bibitem[Finlator et al.(2007)]{2007MNRAS.376.1861F} Finlator, K., 
Dav{\'e}, R., \& Oppenheimer, B.~D.\ 2007, \mnras, 376, 1861 
\bibitem{} Garnett, D. R., Shields, G. A., Skillman, E. D., Sagan, S. P. \& Dufour, R. J. , 1997, ApJ, 489, 63
\bibitem[Hammer et 
al.(2005)]{2005A&A...430..115H} Hammer, F., Flores, H., Elbaz, D., et al.\ 2005, \aap, 430, 115 
\bibitem[Hasinger et al.(2007)]{2007ApJS..172...29H} Hasinger, G., 
Cappelluti, N., Brunner, H., et al.\ 2007, \apjs, 172, 29 
\bibitem{} Henry, R.~B.~C., Edmunds, M.~G., K\"oppen, J.\ 2000, ApJ, 541, 660 
\bibitem{} Ilbert O., et al., 2004, MNRAS, 351, 541 
\bibitem[Kewley et al.(2001)]{2001ApJ...556..121K} Kewley, L.~J., Dopita, 
M.~A., Sutherland, R.~S., Heisler, C.~A., \& Trevena, J.\ 2001, \apj,
\bibitem{} Kewley, L.J. \& Ellison, S.L., 2008, ApJ, 681, 1186, 
\bibitem[Kewley et al.(2006)]{2006MNRAS.372..961K} Kewley, L.~J., Groves, 
B., Kauffmann, G., \& Heckman, T.\ 2006, \mnras, 372, 961 
\bibitem[Kobulnicky et al.(1999)]{1999ApJ...514..544K} Kobulnicky, H.~A., 
Kennicutt, R.~C., Jr., \& Pizagno, J.~L.\ 1999, \apj, 514, 544 
\bibitem[Kobulnicky et al.(2003)]{2003ApJ...599.1006K} Kobulnicky, H.~A., 
Willmer, C.~N.~A., Phillips, A.~C., et al.\ 2003, \apj, 599, 1006 
\bibitem[Koekemoer et al.(2007)]{2007ApJS..172..196K} Koekemoer, A.~M., 
Aussel, H., Calzetti, D., et al.\ 2007, \apjs, 172, 196 
\bibitem[K{\"o}ppen 
\& Hensler(2005)]{2005A&A...434..531K} K{\"o}ppen, J., \& Hensler, G.\ 2005, \aap, 434, 531 
\bibitem[K{\"o}ppen et al.(2007)]{2007MNRAS.375..673K} K{\"o}ppen, J., 
Weidner, C., \& Kroupa, P.\ 2007, \mnras, 375, 673 
\bibitem[] Lamareille, F. 2007, Ph.D.~Thesis
\bibitem[Lamareille et 
al.(2009)]{2009A&A...495...53L} Lamareille, F., Brinchmann, J., Contini, T., et al.\ 2009, \aap, 495, 53 
\bibitem{} Lamareille, F., Contini, T., Le Borgne, J.-F. et al. 2006, A\&A, 417, 839
\bibitem{} Lamareille, F., Mouhcine, M., Contini, T., Lewis, I., \& Maddox, S. 2004, MNRAS, 350, 396
\bibitem[Lara-L{\'o}pez et 
al.(2010)]{2010A&A...521L..53L} Lara-L{\'o}pez, M.~A., Cepa, J., Bongiovanni, A., et al.\ 2010, \aap, 521, L53 
\bibitem[Le F{\`e}vre et al.(2003)]{2003SPIE.4841.1670L} Le F{\`e}vre, O., 
Saisse, M., Mancini, D., et al.\ 2003, \procspie, 4841, 1670 
\bibitem[Lee et al.(2006)]{2006ApJ...647..970L} Lee, H., Skillman, E.~D., 
Cannon, J.~M., et al.\ 2006, \apj, 647, 970 
\bibitem{} Lequeux, J., Peimbert, M., Rayo, J.~F., Serrano, A., \& Torres-Peimbert, S. 1979, A\&A, 80, 155
\bibitem[Liang et 
al.(2004)]{2004A&A...423..867L} Liang, Y.~C., Hammer, F., Flores, H., et al.\ 2004, \aap, 423, 867 
\bibitem[Liang et 
al.(2007)]{2007A&A...474..807L} Liang, Y.~C., Hammer, F., \& Yin, S.~Y.\ 2007, \aap, 474, 807 
\bibitem[Lilly et al.(2009)]{2009ApJS..184..218L} Lilly, S.~J., Le Brun, 
V., Maier, C., et al.\ 2009, \apjs, 184, 218 
\bibitem[Lilly et al.(2007)]{2007ApJS..172...70L} Lilly, S.~J., Le 
F{\`e}vre, O., Renzini, A., et al.\ 2007, \apjs, 172, 70 
\bibitem{} Liu, X., Shapley, A.E., Coil, A.L., Brinchmann, J. \& Ma, C.-P., 2008, ApJ, 678, 758
\bibitem[Maier et al.(2005)]{2005ApJ...634..849M} Maier, C., Lilly, S.~J., 
Carollo, C.~M., Stockton, A., \& Brodwin, M.\ 2005, \apj, 634, 849 
\bibitem[Maier et al.(2006)]{2006ApJ...639..858M} Maier, C., Lilly, S.~J., 
Carollo, C.~M., et al.\ 2006, \apj, 639, 858 
\bibitem[Maier et al.(2004)]{2004A&A...418..475M} Maier, C., Meisenheimer, K., \& Hippelein, H.\ 2004, \aap, 418, 475 
\bibitem[Maiolino et 
al.(2008)]{2008A&A...488..463M} Maiolino, R., Nagao, T., Grazian, A., et al.\ 2008, \aap, 488, 463 
\bibitem[Mannucci et al.(2009)]{2009MNRAS.398.1915M} Mannucci, F., Cresci, 
G., Maiolino, R., et al.\ 2009, \mnras, 398, 1915 
\bibitem[Mannucci et al.(2010)]{2010MNRAS.408.2115M} Mannucci, F., Cresci, 
G., Maiolino, R., Marconi, A., \& Gnerucci, A.\ 2010, \mnras, 408, 2115 
\bibitem[\protect\citeauthoryear{Maraston}{2005}]{2005MNRAS.362..799M} 
Maraston C., 2005, MNRAS, 362, 799 
\bibitem[\protect\citeauthoryear{Marocco, Hache, 
\& Lamareille}{2011}]{2011A&A...531A..71M} Marocco J., Hache E., Lamareille F., 2011, A\&A, 531, A71 
\bibitem[McCracken et al.(2010)]{2010ApJ...708..202M} McCracken, H.~J., 
Capak, P., Salvato, M., et al.\ 2010, \apj, 708, 202 
\bibitem[\protect\citeauthoryear{McGaugh}{1991}]{1991ApJ...380..140M} 
McGaugh S.~S., 1991, ApJ, 380, 140 
\bibitem[Moustakas et al.(2011)]{2011arXiv1112.3300M} Moustakas, J., 
Zaritsky, D., Brown, M., et al.\ 2011, arXiv:1112.3300 
\bibitem{} Nagao, T., Maiolino, R. \& Marconi, A., 2006, A\&A, 459, 85.
\bibitem[Noeske et al.(2007)]{2007ApJ...660L..43N} Noeske, K.~G., Weiner, 
B.~J., Faber, S.~M., et al.\ 2007, \apjl, 660, L43 
\bibitem[\protect\citeauthoryear{Pagel et al.}{1979}]{1979MNRAS.189...95P} 
Pagel B.~E.~J., Edmunds M.~G., Blackwell D.~E., Chun M.~S., Smith G., 1979, 
MNRAS, 189, 95 
\bibitem[Perez et al.(2011)]{2011MNRAS.417..580P} Perez, J., Michel-Dansac, 
L., \& Tissera, P.~B.\ 2011, \mnras, 417, 580 
\bibitem{} P\'erez-Montero, E. \& Contini, T., 2009, MNRAS (PMC09)
\bibitem[P{\'e}rez-Montero et 
al.(2009)]{2009A&A...495...73P} P{\'e}rez-Montero, E., Contini, T., Lamareille, F., et al.\ 2009, \aap, 495, 73 
\bibitem[P{\'e}rez-Montero 
\& D{\'{\i}}az(2005)]{2005MNRAS.361.1063P} P{\'e}rez-Montero, E., \& D{\'{\i}}az, A.~I.\ 2005, \mnras, 361, 1063 
\bibitem{} P\'erez-Montero, E., H\"agele, G.F., Contini, T. \& D\'\i az, A.I. 2007, MNRAS, 381, 125.
\bibitem[Pettini 
\& Pagel(2004)]{2004MNRAS.348L..59P} Pettini, M., \& Pagel, B.~E.~J.\ 2004, \mnras, 348, L59 
\bibitem[Queyrel et 
al.(2009)]{2009A&A...506..681Q} Queyrel, J., Contini, T., P{\'e}rez-Montero, E., et al.\ 2009, \aap, 506, 681 
\bibitem[\protect\citeauthoryear{Queyrel et 
al.}{2012}]{2012A&A...539A..93Q} Queyrel J., et al., 2012, A\&A, 539, A93 
\bibitem{} Pilyugin, L. S. \& Ferrini, F, 2000, A\&A, 358, 72
\bibitem[Pilyugin et 
al.(2003)]{2003A&A...397..487P} Pilyugin, L.~S., Thuan, T.~X., \& V{\'{\i}}lchez, J.~M.\ 2003, A\&A, 397, 487 
\bibitem{} Pozzetti L., et al., 2007, A\&A, 474, 443 
\bibitem[Sakstein et al.(2011)]{2011MNRAS.410.2203S} Sakstein, J., Pipino, 
A., Devriendt, J.~E.~G., \& Maiolino, R.\ 2011, \mnras, 410, 2203 
\bibitem[Sanders et al.(2007)]{2007ApJS..172...86S} Sanders, D.~B., 
Salvato, M., Aussel, H., et al.\ 2007, \apjs, 172, 86 
\bibitem[Savaglio et al.(2005)]{2005ApJ...635..260S} Savaglio, S., 
Glazebrook, K., Le Borgne, D., et al.\ 2005, \apj, 635, 260 
\bibitem[Saviane et 
al.(2008)]{2008A&A...487..901S} Saviane, I., Ivanov, V.~D., Held, E.~V., et al.\ 2008, \aap, 487, 901 
\bibitem[Schinnerer et al.(2007)]{2007ApJS..172...46S} Schinnerer, E., 
Smol{\v c}i{\'c}, V., Carilli, C.~L., et al.\ 2007, \apjs, 172, 46 
\bibitem[Scoville et al.(2007)]{2007ApJS..172....1S} Scoville, N., Aussel, 
H., Brusa, M., et al.\ 2007, \apjs, 172, 1 
\bibitem{} Skillman, E.D., Kennicutt, R. C.  \& Hodge, P.W., 1989, ApJ, 347, 875
\bibitem{} Spergel, D.~N., Verde, L., Peiris, H.~V., et al., 2003, ApJS, 148, 175
\bibitem[Storchi-Bergmann et al.(1994)]{1994ApJ...429..572S} 
Storchi-Bergmann, T., Calzetti, D., \& Kinney, A.~L.\ 1994, \apj, 429, 572 
\bibitem[Storey 
\& Hummer(1995)]{1995MNRAS.272...41S} Storey, P.~J., \& Hummer, D.~G.\ 1995, \mnras, 272, 41 
\bibitem[Taniguchi et al.(2007)]{2007ApJS..172....9T} Taniguchi, Y., 
Scoville, N., Murayama, T., et al.\ 2007, \apjs, 172, 9 
\bibitem{} Thuan, T.~X., Pilyugin, L.~S., \& Zinchenko, I.~A.\ 2010, ApJ, 712, 1029 
\bibitem[Torrey et al.(2012)]{2012ApJ...746..108T} Torrey, P., Cox, T.~J., 
Kewley, L., \& Hernquist, L.\ 2012, \apj, 746, 108 
\bibitem{} Tremonti, C.A., Heckman, T.~M., Kauffmann, G. et al. 2004, ApJ, 613, 898
\bibitem[\protect\citeauthoryear{Yabe et al.}{2012}]{2012PASJ...64...60Y} 
Yabe K., et al., 2012, PASJ, 64, 60 
\bibitem[Yates et al.(2012)]{2012MNRAS.422..215Y} Yates, R.~M., Kauffmann, 
G., \& Guo, Q.\ 2012, \mnras, 422, 215 
\bibitem[Zahid et al.(2011)]{2011ApJ...730..137Z} Zahid, H.~J., Kewley, 
L.~J., \& Bresolin, F.\ 2011, \apj, 730, 137 
\bibitem{} Zaritsky, D., Kennicutt, R.C., \& Huchra, J.P., 1994, ApJ, 420, 87

\end{thebibliography}
\end{document}